\documentclass[11pt, oneside]{article}   	
\usepackage{geometry} 
\usepackage{latexsym}
\usepackage{paralist}
\usepackage{amssymb}
\usepackage{soul}
\usepackage{amsthm}

\usepackage{paralist}

\usepackage{graphics}

\usepackage{amssymb}
\usepackage{soul}

\usepackage{amsthm}

\newcommand{\extd}{{\rm d}}

\theoremstyle{definition}
\newtheorem{example}{Example}%[chapter]

\theoremstyle{plain}
\newtheorem{lemma}[example]{Lemma}

\newtheorem{definition}[example]{Definition}

\newtheorem{assumption}[example]{Assumption}
\newtheorem{guarantee}[example]{Guarantee}

\begin{document}

%\begin{frontmatter}

%% Title, authors and addresses

%% use the tnoteref command within \title for footnotes;
%% use the tnotetext command for theassociated footnote;
%% use the fnref command within \author or \address for footnotes;
%% use the fntext command for theassociated footnote;
%% use the corref command within \author for corresponding author footnotes;
%% use the cortext command for theassociated footnote;
%% use the ead command for the email address,
%% and the form \ead[url] for the home page:
%% \title{Title\tnoteref{label1}}
%% \tnotetext[label1]{}
%% \author{Name\corref{cor1}\fnref{label2}}
%% \ead{email address}
%% \ead[url]{home page}
%% \fntext[label2]{}
%% \cortext[cor1]{}
% \address{Address}
%% \fntext[label3]{}

\date{}

\title{\bf Analogue-digital systems and the modular decomposition of physical behaviour}

\author{\bf Edwin J.\ Beggs and John V.\ Tucker \\ \\ \bf  College of Science \\ \bf Swansea University, Swansea, Wales}

\maketitle

%\address{College of Science,  Swansea University, \\ Singleton Park, Swansea, SA2 8PP, \\ Wales, United Kingdom}

\begin{abstract}
We take a fresh look at analogue-digital systems focussing on their physical behaviour. We model a general analogue-digital system as a physical process controlled by an algorithm by viewing the physical process as physical oracle to the algorithm, generalising the notion of Turing. We develop a theoretical framework for the specification and analysis of such systems that combines five semantical notions: \textit{actual physical behaviour}, \textit{measured behaviour}, \textit{predicted behaviour}, \textit{computed behaviour} and \textit{exceptional behaviour}. Next, we consider the more general and applicable situation of complex processes that exhibit several distinct modes of physical behaviour. Thus, for their design, a \textit{set} of mathematical models may be needed, each model having its own domain of application and representing a particular mode of behaviour or operation of physical reality with its own physical oracle.  The models may be of disparate kinds and, furthermore, not all physical modes may even have a reliable model. We address the questions: \textit{How do we specify algorithms and software that monitor or govern a complex physical situation with many physical modes? How do we specify a portfolio of modes, and the computational problem of transitioning from using one mode to another mode as physical behaviour changes?}  We propose a general definition of an \textit{analogue-digital system with modes}, and show how any diverse set of modes, with or without models, can be bound together, and how the transitions between modes can be determined, by constructing a data type and mode selection functions. We illustrate the ideas of physical modes and our theory by reflecting on simple examples, including driverless racing cars.
\end{abstract}
%
%%\begin{keyword}
%%%% keywords here, in the form: keyword \sep keyword
%%
%%%% PACS codes here, in the form: \PACS code \sep code
%%
%%%% MSC codes here, in the form: \MSC code \sep code
%%%% or \MSC[2008] code \sep code (2000 is the default)
%%
%%\end{keyword}
%
%%\end{frontmatter}
%
%%% \linenumbers
%
%%% main text
%
%

\section{Introduction} \label{vcgaksuy}

A typical analogue-digital system is a system in which a continuous physical component, process or environment is monitored or governed by a discrete algorithmic process. This simple description covers an astonishingly large range of systems from control systems for machines, industrial plant and buildings to systems that monitor and surveil people.\footnote{Hybrid systems, embedded systems and, most recently, cyber-physical systems are types of analogue-digital systems. A cyber-physical system generalises the concept of an embedded system to a network of interacting devices with physical input and output \cite{Henz,Lee}. Cyber-physical systems and the internet of things are central to speculations on the future of manufacturing \cite{INDUSTRIE}.} Autonomous systems -- vehicles, robots, etc -- are made from analogue-digital systems.

To create the software to control an analogue-digital system, the physical must be replaced by an abstract specification that documents certain operations, tests and properties, and constitutes a data type interface between physical quantities and software.  The specification and validation of an analogue-digital system presents certain problems outside software engineering as the reliability of an analogue-digital system depends upon the abstract assumptions made about the physical system. This paper focuses on the design space between the real world system and a formal specification (rather than on implementing a formal specification).
A general desire for autonomy emphasises an important design criterion:
\newline
\newline
\textbf{Autonomous Data Principle.} \textit{The design of the software for an autonomous system should be able to cope with {\em any} data that may be received from, or sent to, the physical components.}
\newline
\newline
We will take a fresh look at some foundational questions about analogue-digital systems and their control. The heart of our analysis of the digital control of a physical system is the dichotomy between the digital world -- which is the only one that the software knows anything about -- and the real world -- which is the only one that actually matters.  This dichotomy with its emphasis on the physical world, together with the Autonomous Data Principle, leads to some interesting questions that we attempt to answer. The first is:

\textit{What are the different data types involved in controlling analogue-digital systems and how do they interact?}

We present a theoretical framework in which five component ideas about data are combined and analysed, namely data associated with 

\medskip
  \begin{compactitem}
    \item actual physical behaviour of a system; 
    \item measurements of actual physical behaviour; 
        \item predictions about behaviour by mathematical models; and 
    \item computations about behaviour using the mathematical models and other rules. 
        \item unexpected behaviour. 
  \end{compactitem}
%
%1. actual physical behaviour of a system; 
%
%2. measurements of actual physical behaviour; 
%
%3. predictions about behaviour by mathematical models; and 
%
%4. computations about behaviour using the mathematical models and other rules. 

\medskip
Complex systems exhibit quite \textit{distinct} kinds of physical behaviour or operation; for example, many machines have start and stop phases and can operate at different rates. These kinds of behaviour are characterised by distinct kinds of data and distinct mathematical models, each model having its own physical domain of application and representing a particular type of operation. The data and models may be of disparate kinds and, certainly, no single mathematical model is adequate.  Furthermore, the models may not cover adequately all the possible types of behaviour, i.e., not all physical modes may have a reliable model. 
Components of complex systems break down.
 These distinct types of behaviour we call \textit{modes}. A second batch of questions is:

\textit{How can modes of physical behaviour, with or without sound mathematical models, be specified independently of software? How are all the modes bound together to make a robust system?  What conditions govern the transition from one mode to another?}

We propose a method of decomposing and partitioning its complex physical behaviour into modes. These modes separate design concerns and can be used to specify 

(1) different physical characteristics, and so may be identified with subsets of phase space; 

(2)  different models of behaviour, e.g.,\ differential equations, neural nets or some form of stochastic evolution;

(3)  exceptional circumstances if none of the mathematical models apply to a mode of behaviour, e.g., monitoring component failure; or 

(4)  different assessments of priorities and decisions, e.g.,\ an air defence system characterising aircraft as friendly, hostile or unknown. 

These four design tasks will shape our theory.
(The design of the software is also based upon logical and algebraic models of the behaviour of the programs which are derived from the semantics of the specification and programming languages employed;  we will not consider this factor.) 

Our method for decomposing physical behaviour leads to a new class of complex analogue-digital systems having distinct modes of operation. Informally, our concept of an analogue-digital system has the form: 
\textit{physical system + interface for data exchange + algorithm}.  Formally, we propose the algorithm treats the physical system as a \textit{physical oracle}, whose queries and responses are mediated by a protocol governing an interface.\footnote{The idea is a physical generalisation of Turing's and Post's ideas about oracles to algorithms in computability theory.}  A natural query to a physical oracle is a request to make a measurement or change a control parameter. A good deal is known about the basic properties of such a physical oracle model in cases where the physical systems are \textit{very} simple and can be faithfully captured by a \textit{single} physical theory and mathematical model  \cite{Oracles1,Oracles2,Axiomatising,ImapctModels,ThreeExperiments}. Here we generalise this concept of an analogue-digital system by introducing modes that employ many physical oracles. To the interface and data exchange protocol are added a set of \textit{mode transition functions} that specify what needs to be done to make a transition from one mode to another.  

%We create a simple geometric description of the system, one which describes both the relationship between the various modes of operation of the system and when transitions between the modes are necessary. Such a geometry needs a topological space and so we construct a simplicial  complex -- a space made of points, lines, triangles, tetrahedra etc., glued together using simple rules -- that we call a \textit{nerve}. The points represent the modes, and the lines or triangles joining them are used to show how well each mode describes the physical system. As this description is very explicit, it forms a data type that can be used represent visually the mode transition protocol in the design of the algorithmic structure of the portfolio. 

The theoretical framework and modes will be introduced in the abstract setting of metric spaces (which are chosen to express error margins). Whilst, in principle, finding a verified control algorithm to solve safety and liveness problems can be done abstractly on the metric space, for applications we commonly need real number coordinates to represent the data that make up states, formulate differential equations to describe the system, and make computations.  We turn from metric spaces to manifolds, where local behaviour is isomorphic to open subsets of some $\mathbb{R}^n$ which provides local coordinates, we develop the theory further under this postulate:

 \textit{The representations of modes of physical behaviour are in 1-1 correspondence with the coordinate charts of the manifold}.

Perhaps it is necessary to emphasise that our aim is to propose a theoretical framework in which the above components find a home and are integrated.  Our main analysis focusses on the physical system and derives ideas for the design of its control system from its possible behaviour. The mathematical framework is a structure upon which different formal methods for specification can build to facilitate a comprehensive software specification. 

%An attempt at exploring the role of modes in design is  \cite{ADmode}. 
      
We are analysing analogue-digital systems from first principles. What we are proposing is complementary to the existing extensive literature on hybrid automata and verification, e.g. \cite{Henz,aeremo}.  There are extensions of the now classic approach that resonate with the compositional concerns in this paper, including hierarchical automata models, see \cite{LeeTriPtol}. Of particular relevance is \cite{MoBiAI} which identifies mode transition with discontinuous changes in physical systems.
      
The paper is in three parts: in Sections 2-4, a general framework built from the five component ideas about data; in Sections 5-6, a method for specifying modes and the transitions between modes and the use of manifolds and charts is given; and, in Section 7, a case study based upon an autonomous car race is developed.

We thank Felix Costa (Lisbon) for many enjoyable and influential conversations on analogue-digital systems; this paper has grown out of our collaboration on  \cite{Oracles1,Oracles2,Axiomatising,ImapctModels,ThreeExperiments}. We also thank Cinzia Giannetti (Swansea) for discussions on digital twins, cyber-physical systems and Industrie 4.0.

\section{Physical systems: general framework} \label{sectdyn}

\subsection{Idealising physical behaviour}

Consider a physical system idealised as a dynamical system with state space $X$ and time $T$. The states $x\in X$ contain physical data that captures certain properties of the physical system, and the behaviour over time of the physical system is represented by a function 
\begin{center}
$f:X\times T\to X$,
\end{center} 
which gives the state  $f(x_0,t)$ of the system at time $t\in T$ starting from initial state  $x_0 \in X$. We write $f(x_0,t)=f_{x_0}(t)$ to model physical behaviour as a path in $X$ starting from $x_0$.
In our theorising, this function models the \textit{actual behaviour} of the physical system in terms of properties used to define the states. The values of $f(x_0,t)$ are exact. 
It is our aim to examine the nature of algorithms that can monitor, predict and control the actual behaviour of such systems. 

As sensors and actuators only work to a finite error, we take both $T$ and $X$ to be metric spaces so that we can set error margins. 
In fact, we take continuous time to be the positive reals $T=[0,\infty) \subset \mathbb{R}$.
We require the system to be able to re-start at arbitrary times, as we shall look at behaviour in stages, re-initialising the initial state. Thus, it will be convenient to be able to start the time evolution at a time $s\ge 0$, rather than  time zero, in state $x_s$, so we also have 
\[
f|_s:X\times [s,\infty)\to X\ 
\]
where we have $f_s(x_s,t)$ equal to the state at time $t\ge s$ for `initial' value $x_s$ at time $t=s$.

However, to control the system,  we assume there is a control space $P$ of physical control parameters, and so further modify 
the time evolution to be a function
\[
f|_s:X\times P\times [s,\infty)\to X\ .
\]
For example, the control space could be the steering direction and gear position for an autonomous vehicle.\footnote{In principle we might have to record both the state and control parameters for past time, as the future evolution might depend on this, but we neglect this use of streams as a distraction to our main concern here.} 

For large systems a single control space could be too complicated; for example, a plane when airborne has not much use for using the brakes on the wheels to control its velocity.  We restrict the control space at different points in the state space. In effect instead of $X\times P$ we have a subset
$C \subset X\times P$. At certain states $x\in X$, only some control options $(x,p)\in X\times P$ exist, and this is most usefully given by a function 

\begin{figure}[htbp]
\begin{center}

\unitlength 0.5 mm
\begin{picture}(110,42)(0,27)
\linethickness{0.3mm}
\put(10,40){\line(1,0){100}}
\linethickness{0.3mm}
\put(70,45){\line(1,0){40}}
\put(70,30){\line(0,1){15}}
\put(110,30){\line(0,1){15}}
\put(70,30){\line(1,0){40}}
\linethickness{0.3mm}
\put(50,65){\line(1,0){25}}
\put(75,35){\line(0,1){30}}
\put(50,35){\line(1,0){25}}
\put(50,35){\line(0,1){30}}
\linethickness{0.3mm}
\put(10,50){\line(1,0){45}}
\put(10,50){\line(0,1){10}}
\put(55,50){\line(0,1){10}}
\put(10,60){\line(1,0){45}}
\put(35,35){\makebox(0,0)[cc]{$X$}}

\end{picture}

\caption{Control spaces shown vertically vs.\ state space horizontally}
\label{figcontrol}
\end{center}
\vspace{-0.15in}
\end{figure}
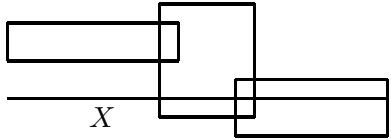

\begin{center}
$\pi:C \to X$ such that the set of control parameters allowed at state $x\in X$ is $\pi^{-1}\{x\}$. 
\end{center}
Then $C=\cup_{x\in X}\pi^{-1}(x)$, and this is pictured in Figure~\ref{figcontrol}, where $X$ is the horizontal line, the control spaces are pictured vertically, $C $ is the union of the rectangles, and $\pi$ is vertical projection to the line. This is similar to the idea of fibration with base $X$.

Now we have 
\[
f|_s:C   \times [s,\infty)\to X\ ,
\]
so for $c_s\in C  $ we write $f|_s(c_s,t)$ for the state at time $t\ge s$ corresponding to the initial state $\pi(c_s)$ and control parameters given by $c_s$ at time $t=s$.

\noindent

\subsection{Models and Measurement} \label{modmes}
We suppose that the actual behaviour of the system in time -- given by a function $f$ as before -- has a mathematical model that specifies a continuous function
\[
x:X\times [0,\infty)\to X
\]
and, in the control case with $C \subset X \times P$,
\[
x:C \times [0,\infty)\to X\ .
\]

The actual behaviour is distinct from the predicted behaviour. For example, $x$ might be a solution to a differential equation, and $x(x_0,t)$ a mathematical prediction that approximates (or fails to approximate) $f(x_0,t)$. Furthermore, it is expected that the predicted behaviour will be distinct from the computed behaviour $y(x_0,t)$, as approximation methods for mathematical equations are not exact.

However, to control a system we have to monitor the behaviour of the system, and for this we need physical measurements. That the actual behaviour is distinct from the measured behaviour is another important distinction.  In theory and practice, two assumptions are employed:

1. measurements are based upon systems units and subunits and, hence, are rational numbers; and

2. measurements have error margins.

\noindent Suppose we have given a space of measurements $M$ and a measurement stream
\[
m_s:C \times [s,\infty)\to M\ .
\]
This has the interpretation that $m_s(c,t)$ is the measurement of the behaviour returned at time $t\ge s$ when the control and state pair is $c=(p,x)\in C$ at time $s\ge 0$. 

Note that the measurements are determined by the control parameter at time $s$; this is hardly surprising since one of the most obvious control messages is to \textit{request} a measurement from the system. We have a time parameter as it may take some time for the result of a request for a measurement to be returned, and also unexpected measurements may occur (e.g., \ fire in an engine). Typically, measurements are taken at regular intervals.

\begin{figure}[htbp]
\begin{center}
\unitlength 0.67 mm
\scalebox{.75}{
\begin{picture}(150,70)(0,13)
\linethickness{0.3mm}
\put(10,70){\line(1,0){30}}
\put(10,20){\line(0,1){50}}
\put(40,20){\line(0,1){50}}
\put(10,20){\line(1,0){30}}
\linethickness{0.3mm}
\put(55,80){\line(1,0){35}}
\put(55,10){\line(0,1){70}}
\put(90,10){\line(0,1){70}}
\put(55,10){\line(1,0){35}}
\linethickness{0.3mm}
\put(105,70){\line(1,0){30}}
\put(105,20){\line(0,1){50}}
\put(135,20){\line(0,1){50}}
\put(105,20){\line(1,0){30}}
\put(25,57){\makebox(0,0)[cc]{physical}}
\put(25,50){\makebox(0,0)[cc]{system}}
\put(25,43){\makebox(0,0)[cc]{$X$}}

\put(72,47){\makebox(0,0)[cc]{interface}} 
\put(72,30){\makebox(0,0)[cc]{control}} %65
\put(72.7,65){\makebox(0,0)[cc]{measurement}} %30

\linethickness{0.3mm}
\multiput(40,60)(0.36,0.12){42}{\line(1,0){0.36}}
\put(55,65){\vector(3,1){0.12}}
\linethickness{0.3mm}
\multiput(40,35)(0.36,-0.12){42}{\line(1,0){0.36}}
\put(40,35){\vector(-3,1){0.12}}
\linethickness{0.3mm}
\multiput(90,65)(0.36,-0.12){42}{\line(1,0){0.36}}
\put(105,60){\vector(3,-1){0.12}}
\linethickness{0.3mm}
\multiput(90,30)(0.36,0.12){42}{\line(1,0){0.36}}
\put(90,30){\vector(-3,-1){0.12}}

\put(120,57){\makebox(0,0)[cc]{algorithm}}
\put(120,50){\makebox(0,0)[cc]{and}}
\put(120,43){\makebox(0,0)[cc]{data}}

\linethickness{0.3mm}
\qbezier(135,55)(142.83,57.65)(146.44,55.84)
\qbezier(146.44,55.84)(150.05,54.04)(150,47.5)
\qbezier(150,47.5)(150.02,40.97)(148.22,38.56)
\qbezier(148.22,38.56)(146.41,36.16)(142.5,37.5)
\qbezier(142.5,37.5)(138.59,38.8)(136.78,39.41)
\qbezier(136.78,39.41)(134.98,40.01)(135,40)
\linethickness{0.3mm}
\multiput(135.62,40)(0.27,-0.12){16}{\line(1,0){0.27}}
\put(135.62,40){\vector(-2,1){0.12}}
\put(146,48.12){\makebox(0,0)[cc]{$y$}}
\put(1,48.12){\makebox(0,0)[cc]{$f$}}

%\put(95,34.38){\makebox(0,0)[cc]{$m$}}

\linethickness{0.3mm}
\qbezier(10,55)(9.77,55.42)(7.07,57.93)
\qbezier(7.07,57.93)(4.38,60.45)(2,60)
\qbezier(2,60)(-2.54,57.74)(-3.34,51.65)
\qbezier(-3.34,51.65)(-4.14,45.56)(-3,40)
\qbezier(-3,40)(-2.58,37.76)(-1.32,35.79)
\qbezier(-1.32,35.79)(-0.05,33.83)(2,33)
\qbezier(2,33)(3.96,32.44)(5.94,33.42)
\qbezier(5.94,33.42)(7.92,34.4)(10,35)
\linethickness{0.3mm}
\multiput(5,33)(0.29,0.12){17}{\line(1,0){0.29}}
\put(10,35){\vector(3,1){0.12}}
\end{picture}
}
\caption{Physical evolution $f$ versus calculated evolution $y$}
\label{phycal}
\end{center}
\vspace{-0.15in}
\end{figure}
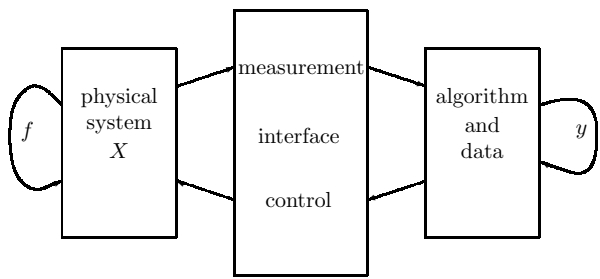

Measurements constitute a \textit{physical oracle} in the sense of \cite{Oracles1,Oracles2,Axiomatising}.
The physical oracle model can encompass control systems. In the case of an aircraft,  oracle queries could request measurements are taken of position, airspeed, fuel amount etc., and also send messages to actuators such as hydraulic rams controlling the wing flaps or wheels.
Thus, these items fit into the standard picture of physical oracles in Figure~\ref{phycal}.

Our main motivation at this stage is to make explicit and distinguish

\medskip\noindent $\bullet$\quad
\textit{actual behaviour} $f$ of the system properties $X$
\newline
$\bullet$\quad \textit{measured behaviour} $m$ valued in $M$
\newline
$\bullet$\quad  \textit{modelled behaviour} $x$ on $X$
\newline
$\bullet$\quad \textit{computed behaviour} $y$, calculated from the model $x$ and measurements $m$.
\medskip

Each of these involve error margins: we expect to calibrate measurements $m$ against actuality $f$ and computation $y$ against prediction $x$. For an analogue-digital system, the calibration of prediction $x$ against actuality $f$ is reduced to the calibration of computation $y$ against measurement $m$. 

How realistic is this abstraction? Of course, we have made a number of simplifications. We have assumed: 

\medskip\noindent 
(i)  a deterministic system, with errors coming from measurement, but the system itself may be significantly stochastic; 
\newline (ii) the mathematical model is reasonable, but it may be inaccurate in certain situations; 
\newline (iii) the infrastructure is functioning, but eventually have mechanical failure of sensors or actuators; 
\newline (iv) the system is independent but may also be subject to control requests from other algorithms, which may even be of hostile intent. 

\medskip
We have derived a partial definition of \textit{digital twin}, an idea of a digital model containing essentially all the data needed to describe the actual system, which was first introduced in manufacturing (for the history and context of the term see \cite{GVtwin}). In this case we have a model of the data and also the dynamics. 

\subsection{Specifications} \label{specyy}
The control of a dynamical system 
\[
f|_s:C\times [s,\infty)\to X\ .
\]
presupposes specifications of what we wish the system to do and what we wish to prevent it from doing. We formalise these properties as follows.

\begin{definition}[Monitoring and control problem] \label{moncon}
Assume we have a \textit{start zone} $S\subset X$, an \textit{end zone} $E\subset X$ and an \textit{avoidance zone} $A\subset X$. By monitoring and controlling the system, we wish to make the system behave as follows: 

(1) starting in any initial state $x_0\in S$ at time $t=0$ the system will arrive eventually, at some time $t_{\mathrm{end}}$ which may depend on $x_0$, in the end zone, i.e.,\ $f(x_0,t_{\mathrm{end}})\in E$. 

(2) The entire behaviour path must be chosen to avoid a $A$, i.e.,\ for all $t\in [0,t_{\mathrm{end}}]$ we have $f(x_0,t)\notin A$. 
\end{definition}

\begin{figure}[htbp]
\begin{center}

\unitlength0.6 mm
\scalebox{.75}{
\begin{picture}(160,67)(0,6)
\linethickness{0.3mm}
\put(70,75){\line(1,0){40}}
\put(70,35){\line(0,1){40}}
\put(110,35){\line(0,1){40}}
\put(70,35){\line(1,0){40}}
\linethickness{0.3mm}
\qbezier(42,72)(46.19,71.01)(46.91,66.32)
\qbezier(46.91,66.32)(47.63,61.63)(45,52.5)
\qbezier(45,52.5)(42.45,43.37)(37.03,39.16)
\qbezier(37.03,39.16)(31.62,34.95)(22.5,35)
\qbezier(22.5,35)(13.38,34.96)(8.56,37.97)
\qbezier(8.56,37.97)(3.75,40.98)(2.5,47.5)
\qbezier(2.5,47.5)(1.17,54.01)(2.26,58.34)
\qbezier(2.26,58.34)(3.34,62.67)(7,65.5)
\qbezier(7,65.5)(10.62,68.37)(14.71,70.05)
\qbezier(14.71,70.05)(18.8,71.73)(24,72.5)
\qbezier(24,72.5)(29.19,73.29)(33.52,73.17)
\qbezier(33.52,73.17)(37.86,73.05)(42,72)
\linethickness{0.3mm}
\qbezier(142.5,72.5)(151.64,71.25)(155.25,66.44)
\qbezier(155.25,66.44)(158.86,61.62)(157.5,52.5)
\qbezier(157.5,52.5)(156.25,43.38)(151.44,38.56)
\qbezier(151.44,38.56)(146.62,33.75)(137.5,32.5)
\qbezier(137.5,32.5)(128.38,31.16)(123.56,32.97)
\qbezier(123.56,32.97)(118.75,34.77)(117.5,40)
\qbezier(117.5,40)(116.18,45.18)(116.78,50.59)
\qbezier(116.78,50.59)(117.38,56.01)(120,62.5)
\qbezier(120,62.5)(122.55,69.03)(127.97,71.44)
\qbezier(127.97,71.44)(133.38,73.84)(142.5,72.5)
\linethickness{0.3mm}
\put(35,49.75){\line(0,1){0.51}}
\multiput(34.98,49.24)(0.02,0.51){1}{\line(0,1){0.51}}
\multiput(34.94,48.74)(0.04,0.5){1}{\line(0,1){0.5}}
\multiput(34.88,48.24)(0.06,0.5){1}{\line(0,1){0.5}}
\multiput(34.81,47.74)(0.08,0.5){1}{\line(0,1){0.5}}
\multiput(34.71,47.25)(0.1,0.49){1}{\line(0,1){0.49}}
\multiput(34.6,46.77)(0.11,0.48){1}{\line(0,1){0.48}}
\multiput(34.46,46.29)(0.13,0.48){1}{\line(0,1){0.48}}
\multiput(34.32,45.82)(0.15,0.47){1}{\line(0,1){0.47}}
\multiput(34.15,45.37)(0.17,0.45){1}{\line(0,1){0.45}}
\multiput(33.96,44.93)(0.09,0.22){2}{\line(0,1){0.22}}
\multiput(33.76,44.5)(0.1,0.21){2}{\line(0,1){0.21}}
\multiput(33.55,44.08)(0.11,0.21){2}{\line(0,1){0.21}}
\multiput(33.31,43.68)(0.12,0.2){2}{\line(0,1){0.2}}
\multiput(33.07,43.3)(0.12,0.19){2}{\line(0,1){0.19}}
\multiput(32.8,42.93)(0.13,0.18){2}{\line(0,1){0.18}}
\multiput(32.53,42.58)(0.14,0.17){2}{\line(0,1){0.17}}
\multiput(32.24,42.25)(0.14,0.17){2}{\line(0,1){0.17}}
\multiput(31.94,41.94)(0.1,0.1){3}{\line(0,1){0.1}}
\multiput(31.63,41.65)(0.16,0.14){2}{\line(1,0){0.16}}
\multiput(31.3,41.38)(0.16,0.13){2}{\line(1,0){0.16}}
\multiput(30.97,41.14)(0.17,0.12){2}{\line(1,0){0.17}}
\multiput(30.63,40.91)(0.17,0.11){2}{\line(1,0){0.17}}
\multiput(30.28,40.71)(0.17,0.1){2}{\line(1,0){0.17}}
\multiput(29.93,40.54)(0.36,0.18){1}{\line(1,0){0.36}}
\multiput(29.56,40.39)(0.36,0.15){1}{\line(1,0){0.36}}
\multiput(29.2,40.26)(0.37,0.13){1}{\line(1,0){0.37}}
\multiput(28.82,40.16)(0.37,0.1){1}{\line(1,0){0.37}}
\multiput(28.45,40.08)(0.38,0.08){1}{\line(1,0){0.38}}
\multiput(28.07,40.03)(0.38,0.05){1}{\line(1,0){0.38}}
\multiput(27.69,40)(0.38,0.03){1}{\line(1,0){0.38}}
\put(27.31,40){\line(1,0){0.38}}
\multiput(26.93,40.03)(0.38,-0.03){1}{\line(1,0){0.38}}
\multiput(26.55,40.08)(0.38,-0.05){1}{\line(1,0){0.38}}
\multiput(26.18,40.16)(0.38,-0.08){1}{\line(1,0){0.38}}
\multiput(25.8,40.26)(0.37,-0.1){1}{\line(1,0){0.37}}
\multiput(25.44,40.39)(0.37,-0.13){1}{\line(1,0){0.37}}
\multiput(25.07,40.54)(0.36,-0.15){1}{\line(1,0){0.36}}
\multiput(24.72,40.71)(0.36,-0.18){1}{\line(1,0){0.36}}
\multiput(24.37,40.91)(0.17,-0.1){2}{\line(1,0){0.17}}
\multiput(24.03,41.14)(0.17,-0.11){2}{\line(1,0){0.17}}
\multiput(23.7,41.38)(0.17,-0.12){2}{\line(1,0){0.17}}
\multiput(23.37,41.65)(0.16,-0.13){2}{\line(1,0){0.16}}
\multiput(23.06,41.94)(0.16,-0.14){2}{\line(1,0){0.16}}
\multiput(22.76,42.25)(0.1,-0.1){3}{\line(0,-1){0.1}}
\multiput(22.47,42.58)(0.14,-0.17){2}{\line(0,-1){0.17}}
\multiput(22.2,42.93)(0.14,-0.17){2}{\line(0,-1){0.17}}
\multiput(21.93,43.3)(0.13,-0.18){2}{\line(0,-1){0.18}}
\multiput(21.69,43.68)(0.12,-0.19){2}{\line(0,-1){0.19}}
\multiput(21.45,44.08)(0.12,-0.2){2}{\line(0,-1){0.2}}
\multiput(21.24,44.5)(0.11,-0.21){2}{\line(0,-1){0.21}}
\multiput(21.04,44.93)(0.1,-0.21){2}{\line(0,-1){0.21}}
\multiput(20.85,45.37)(0.09,-0.22){2}{\line(0,-1){0.22}}
\multiput(20.68,45.82)(0.17,-0.45){1}{\line(0,-1){0.45}}
\multiput(20.54,46.29)(0.15,-0.47){1}{\line(0,-1){0.47}}
\multiput(20.4,46.77)(0.13,-0.48){1}{\line(0,-1){0.48}}
\multiput(20.29,47.25)(0.11,-0.48){1}{\line(0,-1){0.48}}
\multiput(20.19,47.74)(0.1,-0.49){1}{\line(0,-1){0.49}}
\multiput(20.12,48.24)(0.08,-0.5){1}{\line(0,-1){0.5}}
\multiput(20.06,48.74)(0.06,-0.5){1}{\line(0,-1){0.5}}
\multiput(20.02,49.24)(0.04,-0.5){1}{\line(0,-1){0.5}}
\multiput(20,49.75)(0.02,-0.51){1}{\line(0,-1){0.51}}
\put(20,49.75){\line(0,1){0.51}}
\multiput(20,50.25)(0.02,0.51){1}{\line(0,1){0.51}}
\multiput(20.02,50.76)(0.04,0.5){1}{\line(0,1){0.5}}
\multiput(20.06,51.26)(0.06,0.5){1}{\line(0,1){0.5}}
\multiput(20.12,51.76)(0.08,0.5){1}{\line(0,1){0.5}}
\multiput(20.19,52.26)(0.1,0.49){1}{\line(0,1){0.49}}
\multiput(20.29,52.75)(0.11,0.48){1}{\line(0,1){0.48}}
\multiput(20.4,53.23)(0.13,0.48){1}{\line(0,1){0.48}}
\multiput(20.54,53.71)(0.15,0.47){1}{\line(0,1){0.47}}
\multiput(20.68,54.18)(0.17,0.45){1}{\line(0,1){0.45}}
\multiput(20.85,54.63)(0.09,0.22){2}{\line(0,1){0.22}}
\multiput(21.04,55.07)(0.1,0.21){2}{\line(0,1){0.21}}
\multiput(21.24,55.5)(0.11,0.21){2}{\line(0,1){0.21}}
\multiput(21.45,55.92)(0.12,0.2){2}{\line(0,1){0.2}}
\multiput(21.69,56.32)(0.12,0.19){2}{\line(0,1){0.19}}
\multiput(21.93,56.7)(0.13,0.18){2}{\line(0,1){0.18}}
\multiput(22.2,57.07)(0.14,0.17){2}{\line(0,1){0.17}}
\multiput(22.47,57.42)(0.14,0.17){2}{\line(0,1){0.17}}
\multiput(22.76,57.75)(0.1,0.1){3}{\line(0,1){0.1}}
\multiput(23.06,58.06)(0.16,0.14){2}{\line(1,0){0.16}}
\multiput(23.37,58.35)(0.16,0.13){2}{\line(1,0){0.16}}
\multiput(23.7,58.62)(0.17,0.12){2}{\line(1,0){0.17}}
\multiput(24.03,58.86)(0.17,0.11){2}{\line(1,0){0.17}}
\multiput(24.37,59.09)(0.17,0.1){2}{\line(1,0){0.17}}
\multiput(24.72,59.29)(0.36,0.18){1}{\line(1,0){0.36}}
\multiput(25.07,59.46)(0.36,0.15){1}{\line(1,0){0.36}}
\multiput(25.44,59.61)(0.37,0.13){1}{\line(1,0){0.37}}
\multiput(25.8,59.74)(0.37,0.1){1}{\line(1,0){0.37}}
\multiput(26.18,59.84)(0.38,0.08){1}{\line(1,0){0.38}}
\multiput(26.55,59.92)(0.38,0.05){1}{\line(1,0){0.38}}
\multiput(26.93,59.97)(0.38,0.03){1}{\line(1,0){0.38}}
\put(27.31,60){\line(1,0){0.38}}
\multiput(27.69,60)(0.38,-0.03){1}{\line(1,0){0.38}}
\multiput(28.07,59.97)(0.38,-0.05){1}{\line(1,0){0.38}}
\multiput(28.45,59.92)(0.38,-0.08){1}{\line(1,0){0.38}}
\multiput(28.82,59.84)(0.37,-0.1){1}{\line(1,0){0.37}}
\multiput(29.2,59.74)(0.37,-0.13){1}{\line(1,0){0.37}}
\multiput(29.56,59.61)(0.36,-0.15){1}{\line(1,0){0.36}}
\multiput(29.93,59.46)(0.36,-0.18){1}{\line(1,0){0.36}}
\multiput(30.28,59.29)(0.17,-0.1){2}{\line(1,0){0.17}}
\multiput(30.63,59.09)(0.17,-0.11){2}{\line(1,0){0.17}}
\multiput(30.97,58.86)(0.17,-0.12){2}{\line(1,0){0.17}}
\multiput(31.3,58.62)(0.16,-0.13){2}{\line(1,0){0.16}}
\multiput(31.63,58.35)(0.16,-0.14){2}{\line(1,0){0.16}}
\multiput(31.94,58.06)(0.1,-0.1){3}{\line(0,-1){0.1}}
\multiput(32.24,57.75)(0.14,-0.17){2}{\line(0,-1){0.17}}
\multiput(32.53,57.42)(0.14,-0.17){2}{\line(0,-1){0.17}}
\multiput(32.8,57.07)(0.13,-0.18){2}{\line(0,-1){0.18}}
\multiput(33.07,56.7)(0.12,-0.19){2}{\line(0,-1){0.19}}
\multiput(33.31,56.32)(0.12,-0.2){2}{\line(0,-1){0.2}}
\multiput(33.55,55.92)(0.11,-0.21){2}{\line(0,-1){0.21}}
\multiput(33.76,55.5)(0.1,-0.21){2}{\line(0,-1){0.21}}
\multiput(33.96,55.07)(0.09,-0.22){2}{\line(0,-1){0.22}}
\multiput(34.15,54.63)(0.17,-0.45){1}{\line(0,-1){0.45}}
\multiput(34.32,54.18)(0.15,-0.47){1}{\line(0,-1){0.47}}
\multiput(34.46,53.71)(0.13,-0.48){1}{\line(0,-1){0.48}}
\multiput(34.6,53.23)(0.11,-0.48){1}{\line(0,-1){0.48}}
\multiput(34.71,52.75)(0.1,-0.49){1}{\line(0,-1){0.49}}
\multiput(34.81,52.26)(0.08,-0.5){1}{\line(0,-1){0.5}}
\multiput(34.88,51.76)(0.06,-0.5){1}{\line(0,-1){0.5}}
\multiput(34.94,51.26)(0.04,-0.5){1}{\line(0,-1){0.5}}
\multiput(34.98,50.76)(0.02,-0.51){1}{\line(0,-1){0.51}}

\linethickness{0.3mm}
\multiput(147.47,46.5)(0.06,-0.49){1}{\line(0,-1){0.49}}
\multiput(147.53,46)(0.04,-0.49){1}{\line(0,-1){0.49}}
\multiput(147.57,45.51)(0.03,-0.49){1}{\line(0,-1){0.49}}
\multiput(147.59,45.02)(0.01,-0.49){1}{\line(0,-1){0.49}}
\multiput(147.6,44.03)(0.01,0.49){1}{\line(0,1){0.49}}
\multiput(147.57,43.54)(0.02,0.49){1}{\line(0,1){0.49}}
\multiput(147.53,43.05)(0.04,0.49){1}{\line(0,1){0.49}}
\multiput(147.48,42.57)(0.06,0.48){1}{\line(0,1){0.48}}
\multiput(147.41,42.09)(0.07,0.48){1}{\line(0,1){0.48}}
\multiput(147.32,41.62)(0.09,0.47){1}{\line(0,1){0.47}}
\multiput(147.22,41.15)(0.1,0.47){1}{\line(0,1){0.47}}
\multiput(147.1,40.69)(0.12,0.46){1}{\line(0,1){0.46}}
\multiput(146.97,40.25)(0.13,0.45){1}{\line(0,1){0.45}}
\multiput(146.82,39.81)(0.15,0.44){1}{\line(0,1){0.44}}
\multiput(146.66,39.38)(0.16,0.43){1}{\line(0,1){0.43}}
\multiput(146.48,38.97)(0.18,0.41){1}{\line(0,1){0.41}}
\multiput(146.29,38.57)(0.1,0.2){2}{\line(0,1){0.2}}
\multiput(146.08,38.18)(0.1,0.19){2}{\line(0,1){0.19}}
\multiput(145.87,37.81)(0.11,0.19){2}{\line(0,1){0.19}}
\multiput(145.64,37.45)(0.11,0.18){2}{\line(0,1){0.18}}
\multiput(145.39,37.11)(0.12,0.17){2}{\line(0,1){0.17}}
\multiput(145.14,36.78)(0.13,0.16){2}{\line(0,1){0.16}}
\multiput(144.88,36.47)(0.13,0.15){2}{\line(0,1){0.15}}
\multiput(144.6,36.18)(0.14,0.14){2}{\line(0,1){0.14}}
\multiput(144.32,35.91)(0.14,0.14){2}{\line(1,0){0.14}}
\multiput(144.02,35.66)(0.15,0.13){2}{\line(1,0){0.15}}
\multiput(143.72,35.43)(0.15,0.12){2}{\line(1,0){0.15}}
\multiput(143.41,35.22)(0.16,0.11){2}{\line(1,0){0.16}}
\multiput(143.1,35.03)(0.16,0.09){2}{\line(1,0){0.16}}
\multiput(142.77,34.86)(0.32,0.17){1}{\line(1,0){0.32}}
\multiput(142.44,34.71)(0.33,0.15){1}{\line(1,0){0.33}}
\multiput(142.11,34.59)(0.34,0.13){1}{\line(1,0){0.34}}
\multiput(141.77,34.49)(0.34,0.1){1}{\line(1,0){0.34}}
\multiput(141.42,34.41)(0.34,0.08){1}{\line(1,0){0.34}}
\multiput(141.08,34.35)(0.35,0.06){1}{\line(1,0){0.35}}
\multiput(140.73,34.31)(0.35,0.04){1}{\line(1,0){0.35}}
\multiput(140.38,34.3)(0.35,0.01){1}{\line(1,0){0.35}}
\multiput(140.03,34.31)(0.35,-0.01){1}{\line(1,0){0.35}}
\multiput(139.68,34.35)(0.35,-0.03){1}{\line(1,0){0.35}}
\multiput(139.32,34.4)(0.35,-0.06){1}{\line(1,0){0.35}}
\multiput(138.98,34.48)(0.35,-0.08){1}{\line(1,0){0.35}}
\multiput(138.63,34.58)(0.35,-0.1){1}{\line(1,0){0.35}}
\multiput(138.28,34.71)(0.34,-0.12){1}{\line(1,0){0.34}}
\multiput(137.94,34.85)(0.34,-0.15){1}{\line(1,0){0.34}}
\multiput(137.61,35.02)(0.34,-0.17){1}{\line(1,0){0.34}}
\multiput(137.28,35.21)(0.17,-0.09){2}{\line(1,0){0.17}}
\multiput(136.95,35.42)(0.16,-0.1){2}{\line(1,0){0.16}}
\multiput(136.63,35.65)(0.16,-0.11){2}{\line(1,0){0.16}}
\multiput(136.32,35.9)(0.16,-0.12){2}{\line(1,0){0.16}}
\multiput(136.02,36.16)(0.15,-0.13){2}{\line(1,0){0.15}}
\multiput(135.72,36.45)(0.15,-0.14){2}{\line(1,0){0.15}}
\multiput(135.44,36.76)(0.14,-0.15){2}{\line(0,-1){0.15}}
\multiput(135.16,37.08)(0.14,-0.16){2}{\line(0,-1){0.16}}
\multiput(134.9,37.42)(0.13,-0.17){2}{\line(0,-1){0.17}}
\multiput(134.64,37.78)(0.13,-0.18){2}{\line(0,-1){0.18}}
\multiput(134.4,38.15)(0.12,-0.19){2}{\line(0,-1){0.19}}
\multiput(134.17,38.54)(0.12,-0.19){2}{\line(0,-1){0.19}}
\multiput(133.95,38.94)(0.11,-0.2){2}{\line(0,-1){0.2}}
\multiput(133.74,39.35)(0.1,-0.21){2}{\line(0,-1){0.21}}
\multiput(133.55,39.78)(0.1,-0.21){2}{\line(0,-1){0.21}}
\multiput(133.37,40.22)(0.18,-0.44){1}{\line(0,-1){0.44}}
\multiput(133.21,40.66)(0.16,-0.45){1}{\line(0,-1){0.45}}
\multiput(133.06,41.12)(0.15,-0.46){1}{\line(0,-1){0.46}}
\multiput(132.92,41.58)(0.14,-0.46){1}{\line(0,-1){0.46}}
\multiput(132.8,42.06)(0.12,-0.47){1}{\line(0,-1){0.47}}
\multiput(132.69,42.53)(0.11,-0.48){1}{\line(0,-1){0.48}}
\multiput(132.6,43.02)(0.09,-0.48){1}{\line(0,-1){0.48}}
\multiput(132.53,43.5)(0.07,-0.49){1}{\line(0,-1){0.49}}
\multiput(132.47,44)(0.06,-0.49){1}{\line(0,-1){0.49}}
\multiput(132.43,44.49)(0.04,-0.49){1}{\line(0,-1){0.49}}
\multiput(132.41,44.98)(0.03,-0.49){1}{\line(0,-1){0.49}}
\multiput(132.4,45.48)(0.01,-0.49){1}{\line(0,-1){0.49}}
\multiput(132.4,45.48)(0.01,0.49){1}{\line(0,1){0.49}}
\multiput(132.4,45.97)(0.02,0.49){1}{\line(0,1){0.49}}
\multiput(132.43,46.46)(0.04,0.49){1}{\line(0,1){0.49}}
\multiput(132.47,46.95)(0.06,0.48){1}{\line(0,1){0.48}}
\multiput(132.52,47.43)(0.07,0.48){1}{\line(0,1){0.48}}
\multiput(132.59,47.91)(0.09,0.47){1}{\line(0,1){0.47}}
\multiput(132.68,48.38)(0.1,0.47){1}{\line(0,1){0.47}}
\multiput(132.78,48.85)(0.12,0.46){1}{\line(0,1){0.46}}
\multiput(132.9,49.31)(0.13,0.45){1}{\line(0,1){0.45}}
\multiput(133.03,49.75)(0.15,0.44){1}{\line(0,1){0.44}}
\multiput(133.18,50.19)(0.16,0.43){1}{\line(0,1){0.43}}
\multiput(133.34,50.62)(0.18,0.41){1}{\line(0,1){0.41}}
\multiput(133.52,51.03)(0.1,0.2){2}{\line(0,1){0.2}}
\multiput(133.71,51.43)(0.1,0.19){2}{\line(0,1){0.19}}
\multiput(133.92,51.82)(0.11,0.19){2}{\line(0,1){0.19}}
\multiput(134.13,52.19)(0.11,0.18){2}{\line(0,1){0.18}}
\multiput(134.36,52.55)(0.12,0.17){2}{\line(0,1){0.17}}
\multiput(134.61,52.89)(0.13,0.16){2}{\line(0,1){0.16}}
\multiput(134.86,53.22)(0.13,0.15){2}{\line(0,1){0.15}}
\multiput(135.12,53.53)(0.14,0.14){2}{\line(0,1){0.14}}
\multiput(135.4,53.82)(0.14,0.14){2}{\line(1,0){0.14}}
\multiput(135.68,54.09)(0.15,0.13){2}{\line(1,0){0.15}}
\multiput(135.98,54.34)(0.15,0.12){2}{\line(1,0){0.15}}
\multiput(136.28,54.57)(0.16,0.11){2}{\line(1,0){0.16}}
\multiput(136.59,54.78)(0.16,0.09){2}{\line(1,0){0.16}}
\multiput(136.9,54.97)(0.32,0.17){1}{\line(1,0){0.32}}
\multiput(137.23,55.14)(0.33,0.15){1}{\line(1,0){0.33}}
\multiput(137.56,55.29)(0.34,0.13){1}{\line(1,0){0.34}}
\multiput(137.89,55.41)(0.34,0.1){1}{\line(1,0){0.34}}
\multiput(138.23,55.51)(0.34,0.08){1}{\line(1,0){0.34}}
\multiput(138.58,55.59)(0.35,0.06){1}{\line(1,0){0.35}}
\multiput(138.92,55.65)(0.35,0.04){1}{\line(1,0){0.35}}
\multiput(139.27,55.69)(0.35,0.01){1}{\line(1,0){0.35}}
\multiput(139.62,55.7)(0.35,-0.01){1}{\line(1,0){0.35}}
\multiput(139.97,55.69)(0.35,-0.03){1}{\line(1,0){0.35}}
\multiput(140.32,55.65)(0.35,-0.06){1}{\line(1,0){0.35}}
\multiput(140.68,55.6)(0.35,-0.08){1}{\line(1,0){0.35}}
\multiput(141.02,55.52)(0.35,-0.1){1}{\line(1,0){0.35}}
\multiput(141.37,55.42)(0.34,-0.12){1}{\line(1,0){0.34}}
\multiput(141.72,55.29)(0.34,-0.15){1}{\line(1,0){0.34}}
\multiput(142.06,55.15)(0.34,-0.17){1}{\line(1,0){0.34}}
\multiput(142.39,54.98)(0.17,-0.09){2}{\line(1,0){0.17}}
\multiput(142.72,54.79)(0.16,-0.1){2}{\line(1,0){0.16}}
\multiput(143.05,54.58)(0.16,-0.11){2}{\line(1,0){0.16}}
\multiput(143.37,54.35)(0.16,-0.12){2}{\line(1,0){0.16}}
\multiput(143.68,54.1)(0.15,-0.13){2}{\line(1,0){0.15}}
\multiput(143.98,53.84)(0.15,-0.14){2}{\line(1,0){0.15}}
\multiput(144.28,53.55)(0.14,-0.15){2}{\line(0,-1){0.15}}
\multiput(144.56,53.24)(0.14,-0.16){2}{\line(0,-1){0.16}}
\multiput(144.84,52.92)(0.13,-0.17){2}{\line(0,-1){0.17}}
\multiput(145.1,52.58)(0.13,-0.18){2}{\line(0,-1){0.18}}
\multiput(145.36,52.22)(0.12,-0.19){2}{\line(0,-1){0.19}}
\multiput(145.6,51.85)(0.12,-0.19){2}{\line(0,-1){0.19}}
\multiput(145.83,51.46)(0.11,-0.2){2}{\line(0,-1){0.2}}
\multiput(146.05,51.06)(0.1,-0.21){2}{\line(0,-1){0.21}}
\multiput(146.26,50.65)(0.1,-0.21){2}{\line(0,-1){0.21}}
\multiput(146.45,50.22)(0.18,-0.44){1}{\line(0,-1){0.44}}
\multiput(146.63,49.78)(0.16,-0.45){1}{\line(0,-1){0.45}}
\multiput(146.79,49.34)(0.15,-0.46){1}{\line(0,-1){0.46}}
\multiput(146.94,48.88)(0.14,-0.46){1}{\line(0,-1){0.46}}
\multiput(147.08,48.42)(0.12,-0.47){1}{\line(0,-1){0.47}}
\multiput(147.2,47.94)(0.11,-0.48){1}{\line(0,-1){0.48}}
\multiput(147.31,47.47)(0.09,-0.48){1}{\line(0,-1){0.48}}
\multiput(147.4,46.98)(0.07,-0.49){1}{\line(0,-1){0.49}}

\linethickness{0.3mm}
\qbezier(31.25,58.75)(37.03,52.35)(42.73,46.46)
\qbezier(42.73,46.46)(48.42,40.57)(55,35)
\qbezier(55,35)(59.74,30.82)(64.44,26.98)
\qbezier(64.44,26.98)(69.14,23.13)(75,21)
\qbezier(75,21)(81.12,18.96)(87.4,18.79)
\qbezier(87.4,18.79)(93.68,18.62)(100,20)
\qbezier(100,20)(105.65,21.21)(110.67,23.69)
\qbezier(110.67,23.69)(115.69,26.17)(120,30)
\qbezier(120,30)(124.49,34.14)(127.43,39.28)
\qbezier(127.43,39.28)(130.38,44.42)(133,50)
\linethickness{0.3mm}
\qbezier(24,41)(24.64,40.13)(31.43,32.64)
\qbezier(31.43,32.64)(38.21,25.16)(45,20)
\qbezier(45,20)(52.88,14.25)(61.17,9.75)
\qbezier(61.17,9.75)(69.46,5.26)(79,4)
\qbezier(79,4)(91.11,2.66)(102.81,5.23)
\qbezier(102.81,5.23)(114.5,7.8)(125,14)
\qbezier(125,14)(131.53,18.17)(135.88,24.45)
\qbezier(135.88,24.45)(140.22,30.72)(145,37)
\linethickness{0.3mm}
\qbezier(27,50)(27.13,49.68)(29,46.48)
\qbezier(29,46.48)(30.88,43.27)(33,41)
\qbezier(33,41)(35.48,38.79)(38.47,37.47)
\qbezier(38.47,37.47)(41.45,36.15)(44,34)
\qbezier(44,34)(45.83,32.22)(47.13,30.13)
\qbezier(47.13,30.13)(48.44,28.04)(50,26)
\qbezier(50,26)(51.62,23.64)(53.04,20.98)
\qbezier(53.04,20.98)(54.46,18.31)(57,18)
\qbezier(57,18)(58.31,18.19)(58.99,19.9)
\qbezier(58.99,19.9)(59.68,21.61)(61,22)
\qbezier(61,22)(64.21,22.12)(66.76,19.66)
\qbezier(66.76,19.66)(69.32,17.2)(72,15)
\qbezier(72,15)(73.29,13.97)(74.39,12.75)
\qbezier(74.39,12.75)(75.48,11.54)(77,11)
\qbezier(77,11)(79.47,10.41)(81.94,11.07)
\qbezier(81.94,11.07)(84.41,11.73)(87,12)
\qbezier(87,12)(91.65,12.02)(96.14,11.36)
\qbezier(96.14,11.36)(100.62,10.71)(105,12)
\qbezier(105,12)(107.68,13.07)(109.64,15.19)
\qbezier(109.64,15.19)(111.6,17.31)(114,19)
\qbezier(114,19)(116.97,20.74)(120.06,21.91)
\qbezier(120.06,21.91)(123.16,23.08)(126,25)
\qbezier(126,25)(127.19,25.84)(128.2,26.81)
\qbezier(128.2,26.81)(129.21,27.78)(130,29)
\qbezier(130,29)(131.07,30.9)(131.44,33.05)
\qbezier(131.44,33.05)(131.81,35.19)(133,37)
\qbezier(133,37)(134.62,38.96)(136.82,40.15)
\qbezier(136.82,40.15)(139.03,41.34)(141,43)
\put(27,53){\makebox(0,0)[cc]{$x_0$}}

\put(81,15.5){\makebox(0,0)[cc]{$f(t)$}}

\put(90,63){\makebox(0,0)[cc]{Avoidance}}
\put(90,56){\makebox(0,0)[cc]{zone}}
\put(90,49){\makebox(0,0)[cc]{$A$}}

\put(3,30){\makebox(0,0)[cc]{Start zone $S$}}

\put(160,30){\makebox(0,0)[cc]{End zone $E$}}

\put(131,10){\makebox(0,0)[cc]{tube}}

\end{picture}
}

\caption{Getting safely to the end}
\label{figtube}
\end{center}
\vspace{-0.15in}
\end{figure}
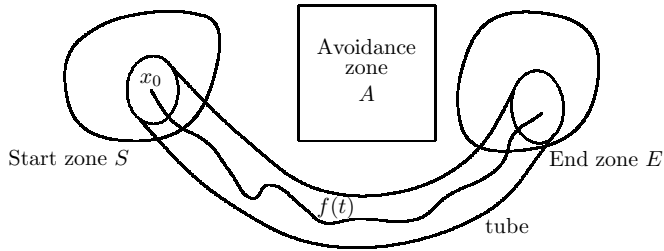

This specification is a combination of a liveness condition (getting to the end) and a safety condition (not hitting the avoidance zone), and
 is illustrated in Figure~\ref{figtube}, with a tube taking account of errors in measurement and control. 
 
 \begin{definition}  \label{tubedef}
 Given a function $y:T\to X$ for some $\eta>0$ we define the $\eta$-{\em{tube}} around $y$ to be
 \[
\big\{(x,t)\in X\times T : d(x,y(t))<\eta\big\}\ .
 \]
 Further we say that $f:T\to X$ lies in the $\eta$-tube if $(f(t),t)$ is in the $\eta$-tube for all $t\in T$.
\end{definition}
 
In the next section we shall generalise this slightly to piecewise paths $y$. A control system will use previous data on the system to alter its behaviour to solve our safety and liveness problem, and we would like to verify that such a system works.

\begin{definition}[A sufficient condition for verification]  \label{etaver}
A method for solving the control problem is {\em{verifiable to precision}} $\eta>0$ if: For any $x_0\in S$, by repeated measurement and control we can construct a path $y:T\to X$, a $t_{\mathrm{end}}\in T$ and an $\eta>0$ so that 
the actual behaviour $f_{x_0}(t)$ lies in the $\eta$-tube around $y(t)$ for $0\le t\le t_{\mathrm{end}}$, the $\eta$-tube does not intersect the avoidance zone, and the $\eta$-tube at time $t_{\mathrm{end}}$ lies within $E$.
\end{definition}

We may not be able to predict the tube in advance, changing circumstances revealed through measurements may force it to be changed in real time.

\section{Computing Behaviour} \label{sectcontrol}
The four components of actual behaviour, modelled behaviour, measurement, and computed behaviour are combined to develop an algorithm for a specification such as in Definition \ref{moncon}.  We give a template or pattern for an algorithm to predict the position of the system in state space, given measured positions as inputs. Without these oracle calls for measurements, the predicted positions would have less and less correspondence with reality. Then we discuss the relevance of this for our specification in Section~\ref{specyy}. Finally, we illustrate the ideas with the example of a space probe.

\subsection{Measurement and prediction} \label{meprco}
We assume measurements are subject to a fixed error margin $\epsilon$ and are sampled at regular intervals, i.e.,\ there is a fixed time step $\lambda>0$ between making the measurements; with these assumptions:

\begin{definition}
Measurement is a physical oracle that when queried at time $n\lambda$ immediately returns the current position of the system in state space within an error of $\epsilon$. Thus, measurement is represented by a map $m:\{0,\lambda,2\lambda,3\lambda,\dots\} \subset T \to X$  that returns a value in $X$ so that $d(m(n\lambda),f(n\lambda))<\epsilon$, where $d$ is the metric on $X$. 
\end{definition}

The zero time delay in returning the values is taken for convenience -- the error in measurement will be enough to deal with for now. 

We show how to construct an $\eta$-tube about a calculated path $y(t)$. Each oracle call for a measurement provides new data to calibrate the construction:  $y(t)$  is therefore a path constructed in stages -- indexed by time intervals of length $\lambda$ --  of the form:
\begin{center}
$y_{n+1}:[n\lambda,(n+1)\lambda]\to X$. 
\end{center}

The aim is that if we start with a measurement with an initial error margin of $\epsilon$ then we will achieve an error margin of $\eta>0$ for the whole time interval, and create a section of tube.  We work in two stages: To begin, we ignore control parameters.  We suppose that we have a method \textbf{calculate path} to calculate a path in $X$. 

\begin{definition}[The $(\lambda,\epsilon,\eta)$-property] \label{vermot}
The method \textbf{calculate path} obeys the $(\lambda,\epsilon,\eta)$-{\em{property}} if for all $n\ge 0$, and any $a\in X$,

1.  the calculated path 
$y_{n+1}:[n\lambda,(n+1)\lambda]\to X$ starts with $y_{n+1}(n\lambda)=a$; and 

2. the actual path $f(t)\in X$ satisfies $d(f(n\lambda),a)<\epsilon$, 

\noindent then
 \[
d( f(t),y_{n+1}(t))<\eta \quad \mathrm{for\ all}\ t\in [n\lambda,(n+1)\lambda]\ . 
 \]
\end{definition}

We now give an algorithm for $y(t)$ so that $d(f(t),y(t))<\eta$ for all $t$ in the interval $[0,n_{\mathrm{max}}\lambda]$, supposing that we have no means of control, we are merely measuring and predicting the position.

\medskip

\textbf{Procedure: measure-predict} 

\textbf{input} $n_{\mathrm{max}}$

\textbf{for} $n=0$ \textbf{to} $n_{\mathrm{max}}-1$ \textbf{do}

\textbf{oracle call} $a:=m(n\lambda)$

\textbf{calculate path} $y_{n+1}:[n\lambda,(n+1)\lambda]\to X$ \textbf{with initial value} $y_{n+1}(n\lambda)=a$

\textbf{rof}

\medskip

\begin{figure}[htbp]
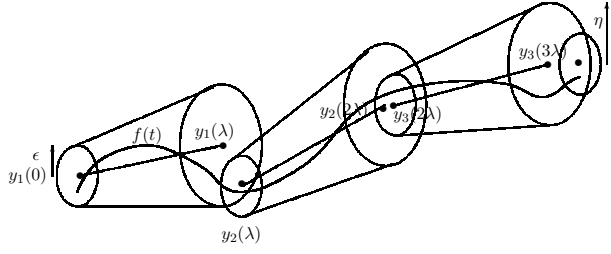

\begin{center}
\unitlength 0.9 mm
\scalebox{.6}{

}
\caption{Measuring and calculating a tube of radius $\eta$}
\label{figdetailtube}
\end{center}
\vspace{-0.15in}
\end{figure}

We piece together the paths $y_{n+1}:[n\lambda,(n+1)\lambda]\to X$ given by this procedure using a disjoint union (which avoids trying to compute with a piecewise continuous function)
\[
y=\bigsqcup_{0\le n <n_{\mathrm{max}}} y_{n+1} :  \bigsqcup_{0\le n <n_{\mathrm{max}}}  [n\lambda,(n+1)\lambda]\to X\ .
\]
The procedure is summed up in Figure~\ref{figdetailtube}, where the actual behaviour of the system is the path $f(t)$; the straight lines connecting the dots are calculated paths; and the ellipses are perspective plots of disks of radius $\epsilon$ and $\eta$ in X. 

\begin{lemma} \label{vermotrec}
Suppose the method \textbf{calculate path} obeys the $(\lambda,\epsilon,\eta)$-property. 
 Then for paths $f(t)$ in the physical system and all $n_{\mathrm{max}}\ge 0$,
 \textbf{Procedure: measure-predict} gives a disjoint union path $y:  \sqcup_{0\le n <n_{\mathrm{max}}}  [n\lambda,(n+1)\lambda]\to X$ so that
 \[
 d(f(t),y(t)) <\eta \ \mathrm{for\ all}\ t\in  \bigsqcup_{0\le n <n_{\mathrm{max}}}  [n\lambda,(n+1)\lambda]\ .
 \]
\end{lemma}

\subsection{Adding control} \label{addcontrol}

Now we introduce control parameters, and in particular we assume that the $(\lambda,\epsilon,\eta)$-property in Definition  ~\ref{vermot} is true where we include a point $b\in C$ of the control and state space with $\pi(b)=a$. 
We assume that the measurements are all of the same nature, so the control parameters will not influence the measurement type, but they still influence the path taken. We modify the initial algorithm pattern as follows:

\medskip
\textbf{Procedure: measure-control-predict} 

\textbf{input} $n_{\mathrm{max}}$

\textbf{for} $n=0$ \textbf{to} $n_{\mathrm{max}}$ \textbf{do}

\textbf{oracle call} $a:=m(n\lambda)$

\textbf{choose} $b\in C $ \textbf{so that}  $\pi(b)=a$

\textbf{calculate path} $y_{n+1}:[n\lambda,(n+1)\lambda]\to X$ \textbf{with initial value} $y_{n+1}(n\lambda)=a$ \newline
$\phantom{f}$ \qquad \textbf{with control} $b$

\textbf{rof}

\medskip
The algorithm calculates a point in the fibre and, taken over all points, a section. It generates two finite streams of measurements $m= m_{1}, \ldots, m_{n_{\mathrm{max}}}$ and control parameters $b= b_{1}, \ldots, b_{n_{\mathrm{max}}}$, where $\pi(b_n) = m(n\lambda)$.

Returning to the idea of \textit{digital twin} mentioned in Section~\ref{modmes}, consider just how well its dynamics reflects the real system. We could say that the digital twin fits the real system to within $(\lambda,\epsilon,\eta)$ if Definition~\ref{vermot} is true. The condition says that we can model the system digitally for a time span $\lambda$, so that errors of size $\epsilon$ do not become bigger than $\eta$ over the time span. This allows both for errors in measurement and some limited stochastic behaviour of the system. Furthermore, to calculate the path in real time, for the result to be useful it must run on the available resources in a small fraction of the time span.

Now, given the notation of \textbf{Procedure: measure-control-predict}, the idea of verification in Definition~\ref{etaver} 
comes down to the existence of controls making a stream b so that the system is verifiable to precision $\eta>0$?
We have the problem of finding an algorithm to generate the stream. If this process is running in real time, then the algorithm to find the control parameters must run \textit{within the time step} $\lambda$. 

\medskip

\textbf{Procedure: measure-control algorithm-predict} 

\textbf{input} $n_{\mathrm{max}}$

\textbf{for} n=0 \textbf{to} $n_{\mathrm{max}}$ \textbf{do}

\textbf{oracle call} $a:=m(n\lambda)$

\textbf{run algorithm to find} $b\in C $ \textbf{so that}  $\pi(b)=a$

\textbf{calculate path} $y_{n+1}:[n\lambda,(n+1)\lambda]\to X$ \textbf{with initial value} $y_{n+1}(n\lambda)=a$ \newline
$\phantom{f}$ \qquad \textbf{with control} $b$

\textbf{rof}

\medskip

\textbf{A tale of two computations:}
It is important to distinguish between two computations, that of (i) the calculated path $y$ and that of (ii) the control parameters $b\in C$. The real time controller for the system has no need to calculate the path at all, provided it is prepared to rely on the algorithm for finding the control parameters that it has been given. In order for this reliance to be reasonable there should be evidence that the control parameter algorithm will actually fulfil the specification in Section~\ref{specyy}. In other words we require evidence that there is an algorithm (calculable in the time step $\lambda$) so that given any initial position $x_0\in S$, so that for any possible outcomes of the measurement oracle (consistent with Proposition~\ref{vermot}) the width $\eta$ tube is guaranteed to avoid the set $A$ and eventually $f(t)$ is in $E$.
Such evidence could come from back of an envelope calculations or from exhaustive empirical observation of the system. However, if we have a digital twin assumed to satisfy the $(\lambda,\epsilon,\eta)$-property then we may attempt to find a verified control algorithm for the specification, by running a supercomputer calculating paths for as long as it takes. We shall consider this further in Section~\ref{versec7}.

\textbf{Taking account of the errors:}
Now that we have explained the procedure, we need to say why in practice it should be slightly modified -- we need to add an extra $+2\epsilon$ to the required clearance from the avoidance zone. The reason is simple: If we have a state a distance $D$ from the avoidance zone, and a measurement with error $\epsilon$ is made, then we only have a measured distance $>D-\epsilon$. If we are relying on measured, rather than calculated, data then according to the measured estimate the actual position could be another $\epsilon$ of the way towards the avoidance zone, i.e.\ distance $>D-2\epsilon$. In other words, the theory may say that we do not enter the avoidance zone, but without the extra $+2\epsilon$ clearance the measured values may be consistent with the system being in the avoidance zone. This would give a clash between measured and predicted states which the system has to resolve, and should be avoided if possible (in real systems this clash may well happen anyway due to `unforeseen circumstances'). For similar reasons we need to take account of an extra $+2\epsilon$ in determining if the system has reached the end region, to avoid a disagreement between the calculated and measured idea of whether the end region has been reached.

\subsection{Example and critique: A space probe}  \label{manpre}
We consider an example where the state space $X$ is an open subset of $\mathbb{R}^n$ with compact closure (i.e.,\ it is a bounded open subset of $\mathbb{R}^n$). The physical system is modelled by a differential equation for $x(t)\in X$ 
\begin{eqnarray} \label{eqrn}
\dot x(t) = v(x(t),t) +  c(x(t),t)  \ .
\end{eqnarray}
We take $v(x,t)$ to be the background vector field modelling the dynamics of the system, and $c(x,t)$ to be the control vector field (this  is the simplest way of applying control to this system). 

\medskip\noindent\textbf{A space probe.}  Imagine a probe in outer space. If we use $(x_1,x_2,x_3)$ to be the spatial coordinates of the probe, to complete the state space we need to have the corresponding velocities $(u_1,u_2,u_3)$ and also the amount of fuel $(F)$, giving the state space as a subset of  $\mathbb{R}^7$. The basic dynamical system is that the probe moves under the influence of gravity, which we denote as an acceleration by the vector
$(g_1,g_2,g_3)$, where the $g_i$ are functions of the position $(x_1,x_2,x_3)$ of the probe in space and time $t$. In addition the probe can use its engine to control its path, and we denote this by the acceleration vector $c=(c_1,c_2,c_3)$. The fuel is depleted at a rate proportional to the magnitude of the acceleration via a constant $k$. Then the equation of motion (\ref{eqrn}) becomes
\begin{eqnarray*}
&&   \big(\frac{\extd x_1}{\extd t},\frac{\extd x_2}{\extd t},\frac{\extd x_3}{\extd t},\frac{\extd u_1}{\extd t},\frac{\extd u_2}{\extd t},\frac{\extd u_3}{\extd t},\frac{\extd F}{\extd t}\big) \cr && \qquad \qquad =\big(u_1,u_2,u_3,g_1,g_2,g_3,0\big) \cr && \qquad  \qquad + \ \big(0,0,0,c_1,c_2,c_3,-k\sqrt{c_1^2+c_2^2+c_3^2}\big).
\end{eqnarray*}
(Specifying the velocities as part of the state space is a standard trick to allow higher order differential equations to be recast as first order ones and thus in the vector field formalism.)

For a space probe the monitoring and control problem of Definition~\ref{moncon}
 becomes specifying 

\noindent (i) an initial position, velocity and fuel allocation (to within an error); 

\noindent (ii) a final position and velocity, and
 
\noindent (iii) the regions to be avoided such as the atmosphere and interior of planets, and running out of fuel, i.e., the region $F<0$.

\medskip\noindent\textbf{Critique of the example.}  
It will be helpful to make some remarks on the limitations of our model for this example.
\newline
\newline
\noindent \textbf{The variation of $\eta$.}
The natural or Euclidean metric on $\mathbb{R}^n$ may not be the one which is best to use for our metric on the state space $X$ (Section ~\ref{sectdyn}). 

A look at the space probe example shows that the coordinates on $\mathbb{R}^7$ are a mix of types -- positions, velocities and fuel amount. At the very least we would have to have scaling factors between these different types, but once we use change of coordinates (as we shall do in Section~\ref{secman}) we shall see that we must allow more general metrics. In general, we shall also have to vary the tolerance of error (called $\eta$ in Section~\ref{sectcontrol}). 

The New Horizons probe to Pluto flew within 12\,500 Km of the surface of Pluto \cite{wikiNH}, and this distance had to be predicted and controlled to considerable accuracy (so $\eta$ is small). However, in the vastness of interplanetary space between Jupiter and Pluto it largely did not matter exactly where the probe was, the tolerance on position was much greater (so $\eta$ is large), and to have it otherwise would be a waste of fuel. 

With this clarification, consider the \textbf{Procedure: measure-control algorithm-predict} from Section~\ref{meprco} for the space probe. The oracle call takes the form of data from Earth tracking stations and possibly cameras on the probe.  To apply the methods of Section~\ref{meprco} to (\ref{eqrn}) we assume that we have a control algorithm to find the control variable $b\in C$ and a differential equation solver to find the path $y_{n+1}$ (e.g.,\ a numerical algorithm such as Runge-Kutta together with a formula for bounding the error). 
\newline
\newline
\noindent\textbf{The value of the time step $\lambda$.}
Suppose that our space probe enters the vicinity of a neutron star, where the gravitational fields vary rapidly over short distances (and therefore short times). 
Even if the computation of the paths may have been handled some time in the past by Earth based supercomputers, the updating of the control parameters given the measurement data must be done in real time, otherwise the probe has no effective control of its trajectory. The measurement methods and on board computer designed for the less demanding environment of the solar system may be unable to cope with the shorter time steps demanded by a rapidly changing environment. 
\newline
\newline
\noindent\textbf{Different trajectories bring different coordinates and strategies.}
Our use of a single coordinate system and equation of motion gives the illusion that any variation of the planned trajectory of the space probe is a minor modification, but this is a long way from the truth. If the trajectory of the probe is altered significantly, so that it visits different objects in the solar system (e.g.,\ involving different gravitational slingshots) then not only do we need entirely different course corrections, but also different schedules of using observing equipment and communications, different mission duration and different assessments of risk. The New Horizons probe had two backup plans (Safe Haven by Other Trajectories, SHBOTs) depending on the probability of collisions with dust, and also after Pluto the mission was altered to visit a Kuiper belt object \cite{wikiNH}. 

Despite their common mathematical structure, the differences in flight trajectory are such that a completely different mission plan is needed for each one. It makes sense to formulate these mission plans separately, but also they inherit a common structure. 

We can interpret these plans as examples of modes in the sense of Section~\ref{secmodone}. In Section~\ref{secman} we shall discuss modes further in an example where even the coordinates required for different modes are different. 

\section{Modes of operation} \label{secmodone}

\subsection{Partitioning the state space and design of modes.}
We examine three stages of a methodology for specifying physical modes, starting from the physical system and the aims for which control is required; the three stages are interconnected. We match them to the four design tasks from Section~\ref{vcgaksuy}.
\newline
\newline
\noindent \textbf{Assigning modes.} 
An analysis and decomposition of behaviour into modes begins the design process, and is before any attempt at software specification. The system to be controlled needs to be modelled, using data that can be measured and laws predicting its evolution (e.g., differential equations). The need for and selection of different models may suggest different modes with different state spaces (tasks 1 \& 2). We also need to consider the \textit{aims}, what the system is designed to do, and this may suggest further sub-division or replication of the previous subsets and models (task 4). The aims of the system are translated into objectives for each mode, and the mathematical model of the mode is used to see what actions by the actuators are required to fulfil the objectives.  For instance, a journey for an aircraft splits into parts that have distinct behaviour -- e.g.,\ taxiing, take off, ascent, cruise, descent, landing, together with various exceptional scenarios such as stall, engine failure --  each of which requires different behaviour to control. 
\newline
\newline
\noindent \textbf{Monitoring infrastructure.} 
The information we need to make decisions largely determines the initial choice of modes but we must consider how the information is obtained and how the decisions reached are implemented. We need to consider what sensors are necessary to obtain the required information, and what actuators are necessary to implement the required control. Devices are likely to function reliably only within given ranges and error margins that further limit certain regions of state space, and possibly give additional constraints on the partition into modes.  For instance, the \textit{instrument landing system} uses sensor systems which are only useable in the vicinity of an airport; an aircraft's rudder is next to useless for manoeuvring while the plane is being pushed back from the gate. 
Further, for safety in case of failure, we may need redundancy -- preferably using different kinds of systems, as systematic failure of a given sensor or actuator type is not unknown (task 3). In the current world environment, we must also take account of the possibility that these devices have been hacked. 
\newline
\newline
\noindent \textbf{Mode transition and objectives.} 
Having established modes, it is vital to be able to change from one mode to another, so we must also determine details of mode transition and how it can be used to implement the intended behaviour of the system and introduce additional safeguards.

%(There is a geometric model of modes and mode transitions as a simplical complex which can be used to visualise this.)

\subsection{The architecture of a multimodal analogue-digital system}\label{Architecture_AD-system}

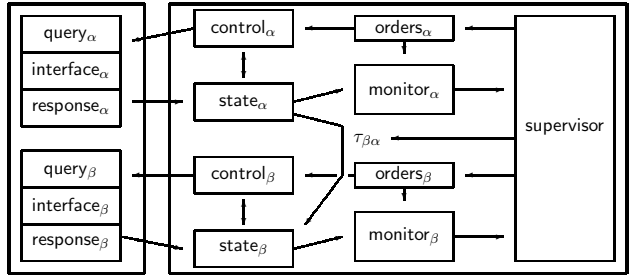
\begin{figure}[htbp]
\begin{center}

\unitlength 0.5 mm
\scalebox{.65}{
\begin{picture}(265,120)(15,-6)
\linethickness{0.3mm}
\put(20,105){\line(1,0){40}}
\put(20,75){\line(0,1){30}}
\put(60,75){\line(0,1){30}}
\put(20,75){\line(1,0){40}}
\linethickness{0.3mm}
\put(20,50){\line(1,0){40}}
\put(20,20){\line(0,1){30}}
\put(60,20){\line(0,1){30}}
\put(20,20){\line(1,0){40}}
\linethickness{0.3mm}
\put(90,107.5){\line(1,0){40}}
\put(90,92.5){\line(0,1){15}}
\put(130,92.5){\line(0,1){15}}
\put(90,92.5){\line(1,0){40}}
\linethickness{0.3mm}
\put(90,77.5){\line(1,0){40}}
\put(90,62.5){\line(0,1){15}}
\put(130,62.5){\line(0,1){15}}
\put(90,62.5){\line(1,0){40}}
\linethickness{0.3mm}
\put(90,47.5){\line(1,0){40}}
\put(90,32.5){\line(0,1){15}}
\put(130,32.5){\line(0,1){15}}
\put(90,32.5){\line(1,0){40}}
\linethickness{0.3mm}
\put(90,17.5){\line(1,0){40}}
\put(90,2.5){\line(0,1){15}}
\put(130,2.5){\line(0,1){15}}
\put(90,2.5){\line(1,0){40}}
\linethickness{0.3mm}
\put(155,85){\line(1,0){40}}
\put(155,65){\line(0,1){20}}
\put(195,65){\line(0,1){20}}
\put(155,65){\line(1,0){40}}
\linethickness{0.3mm}
\put(155,25){\line(1,0){40}}
\put(155,5){\line(0,1){20}}
\put(195,5){\line(0,1){20}}
\put(155,5){\line(1,0){40}}
\linethickness{0.3mm}
\put(220,105){\line(1,0){40}}
\put(220,5){\line(0,1){100}}
\put(260,5){\line(0,1){100}}
\put(220,5){\line(1,0){40}}
\linethickness{0.3mm}
\put(20,90){\line(1,0){40}}
\linethickness{0.3mm}
\put(20,35){\line(1,0){40}}

%%%%%
\put(40,97){\makebox(0,0)[cc]{\textsf{query}${}_\alpha$}}

\put(40,67){\makebox(0,0)[cc]{\textsf{response}${}_\alpha$}}

\put(40,83){\makebox(0,0)[cc]{\textsf{interface}${}_\alpha$}}

\put(110,70){\makebox(0,0)[cc]{\textsf{state}${}_\alpha$}}

\put(110,100){\makebox(0,0)[cc]{\textsf{control}${}_\alpha$}}

\put(175,100){\makebox(0,0)[cc]{\textsf{orders}${}_\alpha$}}

\put(175,75){\makebox(0,0)[cc]{\textsf{monitor}${}_\alpha$}}

\put(40,41){\makebox(0,0)[cc]{\textsf{query}${}_\beta$}}

\put(40,12){\makebox(0,0)[cc]{\textsf{response}${}_\beta$}}

%%%%%%

\put(40,27){\makebox(0,0)[cc]{\textsf{interface}${}_\beta$}}

\put(110,10){\makebox(0,0)[cc]{\textsf{state}${}_\beta$}}

\put(110,40){\makebox(0,0)[cc]{\textsf{control}${}_\beta$}}

\put(175,40){\makebox(0,0)[cc]{\textsf{orders}${}_\beta$}}

\put(175,15){\makebox(0,0)[cc]{\textsf{monitor}${}_\beta$}}

\put(160,54){\makebox(0,0)[cc]{$\tau_{\beta\alpha}$}}

\put(240,60){\makebox(0,0)[cc]{\textsf{supervisor}}}

%\put(135,-14){\makebox(0,0)[cc]{\textbf{Figure 4:} Details of a two mode analogue-digital system}}

%%%%%

\linethickness{0.3mm}
\multiput(65,95)(0.6,0.12){42}{\line(1,0){0.6}}
\put(65,95){\vector(-4,-1){0.12}}
\linethickness{0.3mm}
\put(65,40){\line(1,0){25}}
\put(65,40){\vector(-1,0){0.12}}
\linethickness{0.3mm}
\put(65,70){\line(1,0){20}}
\put(85,70){\vector(1,0){0.12}}
\linethickness{0.3mm}
\multiput(60,15)(0.6,-0.12){42}{\line(1,0){0.6}}
\put(85,10){\vector(4,-1){0.12}}
\linethickness{0.3mm}
\multiput(130,70)(0.48,0.12){42}{\line(1,0){0.48}}
\put(150,75){\vector(4,1){0.12}}
\linethickness{0.3mm}
\multiput(130,10)(0.48,0.12){42}{\line(1,0){0.48}}
\put(150,15){\vector(4,1){0.12}}
\linethickness{0.3mm}
\put(195,75){\line(1,0){20}}
\put(215,75){\vector(1,0){0.12}}
\linethickness{0.3mm}
\put(195,15){\line(1,0){20}}
\put(215,15){\vector(1,0){0.12}}
\linethickness{0.3mm}
\multiput(130,65)(0.48,-0.12){42}{\line(1,0){0.48}}
\linethickness{0.3mm}
\put(150,40){\line(0,1){20}}
\linethickness{0.3mm}
\multiput(135,20)(0.12,0.16){125}{\line(0,1){0.16}}
\put(135,20){\vector(-3,-4){0.12}}

\linethickness{0.3mm}
\put(170,55){\line(1,0){50}}
\put(170,55){\vector(-1,0){0.12}}
\linethickness{0.5mm}
\put(80,110){\line(1,0){185}}
\put(80,0){\line(0,1){110}}
\put(265,0){\line(0,1){110}}
\put(80,0){\line(1,0){185}}
\linethickness{0.5mm}
\put(15,110){\line(1,0){55}}
\put(15,0){\line(0,1){110}}
\put(70,0){\line(0,1){110}}
\put(15,0){\line(1,0){55}}
\linethickness{0.3mm}
\put(110,80){\line(0,1){10}}
\put(110,90){\vector(0,1){0.12}}
\put(110,80){\vector(0,-1){0.12}}
\linethickness{0.3mm}
\put(110,20){\line(0,1){10}}
\put(110,30){\vector(0,1){0.12}}
\put(110,20){\vector(0,-1){0.12}}
\linethickness{0.3mm}
\put(155,105){\line(1,0){40}}
\put(155,95){\line(0,1){10}}
\put(195,95){\line(0,1){10}}
\put(155,95){\line(1,0){40}}
\linethickness{0.3mm}
\put(155,45){\line(1,0){40}}
\put(155,35){\line(0,1){10}}
\put(195,35){\line(0,1){10}}
\put(155,35){\line(1,0){40}}
\linethickness{0.3mm}
\put(200,100){\line(1,0){20}}
\put(200,100){\vector(-1,0){0.12}}
\linethickness{0.3mm}
\put(200,40){\line(1,0){20}}
\put(200,40){\vector(-1,0){0.12}}
\linethickness{0.3mm}
\put(135,100){\line(1,0){20}}
\put(135,100){\vector(-1,0){0.12}}
\linethickness{0.3mm}
\put(135,40){\line(1,0){12.5}}
\put(135,40){\vector(-1,0){0.12}}
\linethickness{0.3mm}
\put(175,90){\line(0,1){5}}
\put(175,90){\vector(0,-1){0.12}}
\linethickness{0.3mm}
\put(175,30){\line(0,1){5}}
\put(175,30){\vector(0,-1){0.12}}
\linethickness{0.3mm}
\put(152.5,40){\line(1,0){2.5}}
\linethickness{0.3mm}
\put(20,5){\line(0,1){15}}
\linethickness{0.3mm}
\put(20,5){\line(1,0){40}}
\linethickness{0.3mm}
\put(60,5){\line(0,1){15}}
\linethickness{0.3mm}
\put(20,60){\line(0,1){15}}
\linethickness{0.3mm}
\put(20,60){\line(1,0){40}}
\linethickness{0.3mm}
\put(60,60){\line(0,1){15}}
\end{picture}
}
\caption{Architecture of a two mode analogue-digital system}
\label{archtwomode}
\end{center}
\vspace{-0.15in}
\end{figure}

 To make the distinction between the physical and the computational worlds it will be useful to consider a possible architecture for a analogue-digital system with only two modes $\alpha$ and $\beta$ and one mode transition $\tau_{\beta\alpha}$ in Figure~\ref{archtwomode}. The hypothetical architecture brings us closer to a specification for software.

The world according to mode $\alpha$ is a collection of digital data $\mathsf{state}_\alpha$, containing data from sensors, calculated data, and constants (= information about the world that is not measured). 
Information from sensors, or action by actuators, is requested by $\mathsf{control}_\alpha$
but what information arrives back, and when, is not under its control. 
Using the information gathered it tries to deduce the actual behaviour of the system, which may mean disregarding information as inaccurate or compromised. Communication with the real world is via $\mathsf{interface}_\alpha$, which relays queries from the control algorithm $\mathsf{control}_\alpha$ to sensors and actuators, and inserts the responses into $\mathsf{state}_\alpha$.  

To complete the architecture, $\mathsf{orders}_\alpha$ contains information related to the objectives of the control system in mode $\alpha$; $\mathsf{monitor}_\alpha$ reviews how well the system is performing related to the objectives in mode $\alpha$; the $\mathsf{supervisor}$ monitors the overall performance of the system, and can modify the objectives in $\mathsf{orders}_\alpha$ or initiate a mode transition from $\alpha$ to $\beta$, which includes a map $\tau_{\beta\alpha}$ transforming data of $\mathsf{state}_\alpha$ to $\mathsf{state}_\beta$.

\subsection{Example of modes: Autonomous boats}
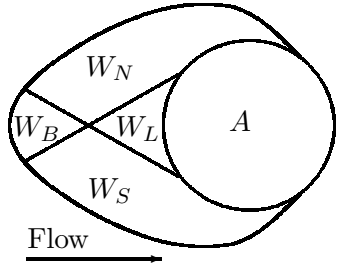
\begin{figure}[htbp]
\begin{center}
\unitlength 0.45 mm
\begin{picture}(100,75)(0,20)
\linethickness{0.3mm}
\put(100,54.75){\line(0,1){0.5}}
\multiput(99.99,54.25)(0.01,0.5){1}{\line(0,1){0.5}}
\multiput(99.97,53.76)(0.02,0.5){1}{\line(0,1){0.5}}
\multiput(99.94,53.26)(0.03,0.5){1}{\line(0,1){0.5}}
\multiput(99.9,52.77)(0.04,0.5){1}{\line(0,1){0.5}}
\multiput(99.85,52.27)(0.05,0.49){1}{\line(0,1){0.49}}
\multiput(99.79,51.78)(0.06,0.49){1}{\line(0,1){0.49}}
\multiput(99.72,51.29)(0.07,0.49){1}{\line(0,1){0.49}}
\multiput(99.64,50.79)(0.08,0.49){1}{\line(0,1){0.49}}
\multiput(99.56,50.31)(0.09,0.49){1}{\line(0,1){0.49}}
\multiput(99.46,49.82)(0.1,0.49){1}{\line(0,1){0.49}}
\multiput(99.35,49.33)(0.11,0.49){1}{\line(0,1){0.49}}
\multiput(99.23,48.85)(0.12,0.48){1}{\line(0,1){0.48}}
\multiput(99.1,48.37)(0.13,0.48){1}{\line(0,1){0.48}}
\multiput(98.97,47.89)(0.14,0.48){1}{\line(0,1){0.48}}
\multiput(98.82,47.42)(0.15,0.48){1}{\line(0,1){0.48}}
\multiput(98.67,46.94)(0.16,0.47){1}{\line(0,1){0.47}}
\multiput(98.5,46.48)(0.16,0.47){1}{\line(0,1){0.47}}
\multiput(98.33,46.01)(0.17,0.47){1}{\line(0,1){0.47}}
\multiput(98.14,45.55)(0.09,0.23){2}{\line(0,1){0.23}}
\multiput(97.95,45.09)(0.1,0.23){2}{\line(0,1){0.23}}
\multiput(97.75,44.64)(0.1,0.23){2}{\line(0,1){0.23}}
\multiput(97.54,44.18)(0.11,0.23){2}{\line(0,1){0.23}}
\multiput(97.32,43.74)(0.11,0.22){2}{\line(0,1){0.22}}
\multiput(97.09,43.3)(0.11,0.22){2}{\line(0,1){0.22}}
\multiput(96.85,42.86)(0.12,0.22){2}{\line(0,1){0.22}}
\multiput(96.61,42.43)(0.12,0.22){2}{\line(0,1){0.22}}
\multiput(96.35,42)(0.13,0.21){2}{\line(0,1){0.21}}
\multiput(96.09,41.58)(0.13,0.21){2}{\line(0,1){0.21}}
\multiput(95.82,41.16)(0.14,0.21){2}{\line(0,1){0.21}}
\multiput(95.54,40.75)(0.14,0.21){2}{\line(0,1){0.21}}
\multiput(95.25,40.35)(0.14,0.2){2}{\line(0,1){0.2}}
\multiput(94.96,39.95)(0.15,0.2){2}{\line(0,1){0.2}}
\multiput(94.66,39.55)(0.1,0.13){3}{\line(0,1){0.13}}
\multiput(94.34,39.16)(0.1,0.13){3}{\line(0,1){0.13}}
\multiput(94.03,38.78)(0.11,0.13){3}{\line(0,1){0.13}}
\multiput(93.7,38.41)(0.11,0.13){3}{\line(0,1){0.13}}
\multiput(93.37,38.04)(0.11,0.12){3}{\line(0,1){0.12}}
\multiput(93.03,37.68)(0.11,0.12){3}{\line(0,1){0.12}}
\multiput(92.68,37.32)(0.12,0.12){3}{\line(0,1){0.12}}
\multiput(92.32,36.97)(0.12,0.12){3}{\line(1,0){0.12}}
\multiput(91.96,36.63)(0.12,0.11){3}{\line(1,0){0.12}}
\multiput(91.59,36.3)(0.12,0.11){3}{\line(1,0){0.12}}
\multiput(91.22,35.97)(0.13,0.11){3}{\line(1,0){0.13}}
\multiput(90.84,35.66)(0.13,0.11){3}{\line(1,0){0.13}}
\multiput(90.45,35.34)(0.13,0.1){3}{\line(1,0){0.13}}
\multiput(90.05,35.04)(0.13,0.1){3}{\line(1,0){0.13}}
\multiput(89.65,34.75)(0.2,0.15){2}{\line(1,0){0.2}}
\multiput(89.25,34.46)(0.2,0.14){2}{\line(1,0){0.2}}
\multiput(88.84,34.18)(0.21,0.14){2}{\line(1,0){0.21}}
\multiput(88.42,33.91)(0.21,0.14){2}{\line(1,0){0.21}}
\multiput(88,33.65)(0.21,0.13){2}{\line(1,0){0.21}}
\multiput(87.57,33.39)(0.21,0.13){2}{\line(1,0){0.21}}
\multiput(87.14,33.15)(0.22,0.12){2}{\line(1,0){0.22}}
\multiput(86.7,32.91)(0.22,0.12){2}{\line(1,0){0.22}}
\multiput(86.26,32.68)(0.22,0.11){2}{\line(1,0){0.22}}
\multiput(85.82,32.46)(0.22,0.11){2}{\line(1,0){0.22}}
\multiput(85.36,32.25)(0.23,0.11){2}{\line(1,0){0.23}}
\multiput(84.91,32.05)(0.23,0.1){2}{\line(1,0){0.23}}
\multiput(84.45,31.86)(0.23,0.1){2}{\line(1,0){0.23}}
\multiput(83.99,31.67)(0.23,0.09){2}{\line(1,0){0.23}}
\multiput(83.52,31.5)(0.47,0.17){1}{\line(1,0){0.47}}
\multiput(83.06,31.33)(0.47,0.16){1}{\line(1,0){0.47}}
\multiput(82.58,31.18)(0.47,0.16){1}{\line(1,0){0.47}}
\multiput(82.11,31.03)(0.48,0.15){1}{\line(1,0){0.48}}
\multiput(81.63,30.9)(0.48,0.14){1}{\line(1,0){0.48}}
\multiput(81.15,30.77)(0.48,0.13){1}{\line(1,0){0.48}}
\multiput(80.67,30.65)(0.48,0.12){1}{\line(1,0){0.48}}
\multiput(80.18,30.54)(0.49,0.11){1}{\line(1,0){0.49}}
\multiput(79.69,30.44)(0.49,0.1){1}{\line(1,0){0.49}}
\multiput(79.21,30.36)(0.49,0.09){1}{\line(1,0){0.49}}
\multiput(78.71,30.28)(0.49,0.08){1}{\line(1,0){0.49}}
\multiput(78.22,30.21)(0.49,0.07){1}{\line(1,0){0.49}}
\multiput(77.73,30.15)(0.49,0.06){1}{\line(1,0){0.49}}
\multiput(77.23,30.1)(0.49,0.05){1}{\line(1,0){0.49}}
\multiput(76.74,30.06)(0.5,0.04){1}{\line(1,0){0.5}}
\multiput(76.24,30.03)(0.5,0.03){1}{\line(1,0){0.5}}
\multiput(75.75,30.01)(0.5,0.02){1}{\line(1,0){0.5}}
\multiput(75.25,30)(0.5,0.01){1}{\line(1,0){0.5}}
\put(74.75,30){\line(1,0){0.5}}
\multiput(74.25,30.01)(0.5,-0.01){1}{\line(1,0){0.5}}
\multiput(73.76,30.03)(0.5,-0.02){1}{\line(1,0){0.5}}
\multiput(73.26,30.06)(0.5,-0.03){1}{\line(1,0){0.5}}
\multiput(72.77,30.1)(0.5,-0.04){1}{\line(1,0){0.5}}
\multiput(72.27,30.15)(0.49,-0.05){1}{\line(1,0){0.49}}
\multiput(71.78,30.21)(0.49,-0.06){1}{\line(1,0){0.49}}
\multiput(71.29,30.28)(0.49,-0.07){1}{\line(1,0){0.49}}
\multiput(70.79,30.36)(0.49,-0.08){1}{\line(1,0){0.49}}
\multiput(70.31,30.44)(0.49,-0.09){1}{\line(1,0){0.49}}
\multiput(69.82,30.54)(0.49,-0.1){1}{\line(1,0){0.49}}
\multiput(69.33,30.65)(0.49,-0.11){1}{\line(1,0){0.49}}
\multiput(68.85,30.77)(0.48,-0.12){1}{\line(1,0){0.48}}
\multiput(68.37,30.9)(0.48,-0.13){1}{\line(1,0){0.48}}
\multiput(67.89,31.03)(0.48,-0.14){1}{\line(1,0){0.48}}
\multiput(67.42,31.18)(0.48,-0.15){1}{\line(1,0){0.48}}
\multiput(66.94,31.33)(0.47,-0.16){1}{\line(1,0){0.47}}
\multiput(66.48,31.5)(0.47,-0.16){1}{\line(1,0){0.47}}
\multiput(66.01,31.67)(0.47,-0.17){1}{\line(1,0){0.47}}
\multiput(65.55,31.86)(0.23,-0.09){2}{\line(1,0){0.23}}
\multiput(65.09,32.05)(0.23,-0.1){2}{\line(1,0){0.23}}
\multiput(64.64,32.25)(0.23,-0.1){2}{\line(1,0){0.23}}
\multiput(64.18,32.46)(0.23,-0.11){2}{\line(1,0){0.23}}
\multiput(63.74,32.68)(0.22,-0.11){2}{\line(1,0){0.22}}
\multiput(63.3,32.91)(0.22,-0.11){2}{\line(1,0){0.22}}
\multiput(62.86,33.15)(0.22,-0.12){2}{\line(1,0){0.22}}
\multiput(62.43,33.39)(0.22,-0.12){2}{\line(1,0){0.22}}
\multiput(62,33.65)(0.21,-0.13){2}{\line(1,0){0.21}}
\multiput(61.58,33.91)(0.21,-0.13){2}{\line(1,0){0.21}}
\multiput(61.16,34.18)(0.21,-0.14){2}{\line(1,0){0.21}}
\multiput(60.75,34.46)(0.21,-0.14){2}{\line(1,0){0.21}}
\multiput(60.35,34.75)(0.2,-0.14){2}{\line(1,0){0.2}}
\multiput(59.95,35.04)(0.2,-0.15){2}{\line(1,0){0.2}}
\multiput(59.55,35.34)(0.13,-0.1){3}{\line(1,0){0.13}}
\multiput(59.16,35.66)(0.13,-0.1){3}{\line(1,0){0.13}}
\multiput(58.78,35.97)(0.13,-0.11){3}{\line(1,0){0.13}}
\multiput(58.41,36.3)(0.13,-0.11){3}{\line(1,0){0.13}}
\multiput(58.04,36.63)(0.12,-0.11){3}{\line(1,0){0.12}}
\multiput(57.68,36.97)(0.12,-0.11){3}{\line(1,0){0.12}}
\multiput(57.32,37.32)(0.12,-0.12){3}{\line(1,0){0.12}}
\multiput(56.97,37.68)(0.12,-0.12){3}{\line(0,-1){0.12}}
\multiput(56.63,38.04)(0.11,-0.12){3}{\line(0,-1){0.12}}
\multiput(56.3,38.41)(0.11,-0.12){3}{\line(0,-1){0.12}}
\multiput(55.97,38.78)(0.11,-0.13){3}{\line(0,-1){0.13}}
\multiput(55.66,39.16)(0.11,-0.13){3}{\line(0,-1){0.13}}
\multiput(55.34,39.55)(0.1,-0.13){3}{\line(0,-1){0.13}}
\multiput(55.04,39.95)(0.1,-0.13){3}{\line(0,-1){0.13}}
\multiput(54.75,40.35)(0.15,-0.2){2}{\line(0,-1){0.2}}
\multiput(54.46,40.75)(0.14,-0.2){2}{\line(0,-1){0.2}}
\multiput(54.18,41.16)(0.14,-0.21){2}{\line(0,-1){0.21}}
\multiput(53.91,41.58)(0.14,-0.21){2}{\line(0,-1){0.21}}
\multiput(53.65,42)(0.13,-0.21){2}{\line(0,-1){0.21}}
\multiput(53.39,42.43)(0.13,-0.21){2}{\line(0,-1){0.21}}
\multiput(53.15,42.86)(0.12,-0.22){2}{\line(0,-1){0.22}}
\multiput(52.91,43.3)(0.12,-0.22){2}{\line(0,-1){0.22}}
\multiput(52.68,43.74)(0.11,-0.22){2}{\line(0,-1){0.22}}
\multiput(52.46,44.18)(0.11,-0.22){2}{\line(0,-1){0.22}}
\multiput(52.25,44.64)(0.11,-0.23){2}{\line(0,-1){0.23}}
\multiput(52.05,45.09)(0.1,-0.23){2}{\line(0,-1){0.23}}
\multiput(51.86,45.55)(0.1,-0.23){2}{\line(0,-1){0.23}}
\multiput(51.67,46.01)(0.09,-0.23){2}{\line(0,-1){0.23}}
\multiput(51.5,46.48)(0.17,-0.47){1}{\line(0,-1){0.47}}
\multiput(51.33,46.94)(0.16,-0.47){1}{\line(0,-1){0.47}}
\multiput(51.18,47.42)(0.16,-0.47){1}{\line(0,-1){0.47}}
\multiput(51.03,47.89)(0.15,-0.48){1}{\line(0,-1){0.48}}
\multiput(50.9,48.37)(0.14,-0.48){1}{\line(0,-1){0.48}}
\multiput(50.77,48.85)(0.13,-0.48){1}{\line(0,-1){0.48}}
\multiput(50.65,49.33)(0.12,-0.48){1}{\line(0,-1){0.48}}
\multiput(50.54,49.82)(0.11,-0.49){1}{\line(0,-1){0.49}}
\multiput(50.44,50.31)(0.1,-0.49){1}{\line(0,-1){0.49}}
\multiput(50.36,50.79)(0.09,-0.49){1}{\line(0,-1){0.49}}
\multiput(50.28,51.29)(0.08,-0.49){1}{\line(0,-1){0.49}}
\multiput(50.21,51.78)(0.07,-0.49){1}{\line(0,-1){0.49}}
\multiput(50.15,52.27)(0.06,-0.49){1}{\line(0,-1){0.49}}
\multiput(50.1,52.77)(0.05,-0.49){1}{\line(0,-1){0.49}}
\multiput(50.06,53.26)(0.04,-0.5){1}{\line(0,-1){0.5}}
\multiput(50.03,53.76)(0.03,-0.5){1}{\line(0,-1){0.5}}
\multiput(50.01,54.25)(0.02,-0.5){1}{\line(0,-1){0.5}}
\multiput(50,54.75)(0.01,-0.5){1}{\line(0,-1){0.5}}
\put(50,54.75){\line(0,1){0.5}}
\multiput(50,55.25)(0.01,0.5){1}{\line(0,1){0.5}}
\multiput(50.01,55.75)(0.02,0.5){1}{\line(0,1){0.5}}
\multiput(50.03,56.24)(0.03,0.5){1}{\line(0,1){0.5}}
\multiput(50.06,56.74)(0.04,0.5){1}{\line(0,1){0.5}}
\multiput(50.1,57.23)(0.05,0.49){1}{\line(0,1){0.49}}
\multiput(50.15,57.73)(0.06,0.49){1}{\line(0,1){0.49}}
\multiput(50.21,58.22)(0.07,0.49){1}{\line(0,1){0.49}}
\multiput(50.28,58.71)(0.08,0.49){1}{\line(0,1){0.49}}
\multiput(50.36,59.21)(0.09,0.49){1}{\line(0,1){0.49}}
\multiput(50.44,59.69)(0.1,0.49){1}{\line(0,1){0.49}}
\multiput(50.54,60.18)(0.11,0.49){1}{\line(0,1){0.49}}
\multiput(50.65,60.67)(0.12,0.48){1}{\line(0,1){0.48}}
\multiput(50.77,61.15)(0.13,0.48){1}{\line(0,1){0.48}}
\multiput(50.9,61.63)(0.14,0.48){1}{\line(0,1){0.48}}
\multiput(51.03,62.11)(0.15,0.48){1}{\line(0,1){0.48}}
\multiput(51.18,62.58)(0.16,0.47){1}{\line(0,1){0.47}}
\multiput(51.33,63.06)(0.16,0.47){1}{\line(0,1){0.47}}
\multiput(51.5,63.52)(0.17,0.47){1}{\line(0,1){0.47}}
\multiput(51.67,63.99)(0.09,0.23){2}{\line(0,1){0.23}}
\multiput(51.86,64.45)(0.1,0.23){2}{\line(0,1){0.23}}
\multiput(52.05,64.91)(0.1,0.23){2}{\line(0,1){0.23}}
\multiput(52.25,65.36)(0.11,0.23){2}{\line(0,1){0.23}}
\multiput(52.46,65.82)(0.11,0.22){2}{\line(0,1){0.22}}
\multiput(52.68,66.26)(0.11,0.22){2}{\line(0,1){0.22}}
\multiput(52.91,66.7)(0.12,0.22){2}{\line(0,1){0.22}}
\multiput(53.15,67.14)(0.12,0.22){2}{\line(0,1){0.22}}
\multiput(53.39,67.57)(0.13,0.21){2}{\line(0,1){0.21}}
\multiput(53.65,68)(0.13,0.21){2}{\line(0,1){0.21}}
\multiput(53.91,68.42)(0.14,0.21){2}{\line(0,1){0.21}}
\multiput(54.18,68.84)(0.14,0.21){2}{\line(0,1){0.21}}
\multiput(54.46,69.25)(0.14,0.2){2}{\line(0,1){0.2}}
\multiput(54.75,69.65)(0.15,0.2){2}{\line(0,1){0.2}}
\multiput(55.04,70.05)(0.1,0.13){3}{\line(0,1){0.13}}
\multiput(55.34,70.45)(0.1,0.13){3}{\line(0,1){0.13}}
\multiput(55.66,70.84)(0.11,0.13){3}{\line(0,1){0.13}}
\multiput(55.97,71.22)(0.11,0.13){3}{\line(0,1){0.13}}
\multiput(56.3,71.59)(0.11,0.12){3}{\line(0,1){0.12}}
\multiput(56.63,71.96)(0.11,0.12){3}{\line(0,1){0.12}}
\multiput(56.97,72.32)(0.12,0.12){3}{\line(0,1){0.12}}
\multiput(57.32,72.68)(0.12,0.12){3}{\line(1,0){0.12}}
\multiput(57.68,73.03)(0.12,0.11){3}{\line(1,0){0.12}}
\multiput(58.04,73.37)(0.12,0.11){3}{\line(1,0){0.12}}
\multiput(58.41,73.7)(0.13,0.11){3}{\line(1,0){0.13}}
\multiput(58.78,74.03)(0.13,0.11){3}{\line(1,0){0.13}}
\multiput(59.16,74.34)(0.13,0.1){3}{\line(1,0){0.13}}
\multiput(59.55,74.66)(0.13,0.1){3}{\line(1,0){0.13}}
\multiput(59.95,74.96)(0.2,0.15){2}{\line(1,0){0.2}}
\multiput(60.35,75.25)(0.2,0.14){2}{\line(1,0){0.2}}
\multiput(60.75,75.54)(0.21,0.14){2}{\line(1,0){0.21}}
\multiput(61.16,75.82)(0.21,0.14){2}{\line(1,0){0.21}}
\multiput(61.58,76.09)(0.21,0.13){2}{\line(1,0){0.21}}
\multiput(62,76.35)(0.21,0.13){2}{\line(1,0){0.21}}
\multiput(62.43,76.61)(0.22,0.12){2}{\line(1,0){0.22}}
\multiput(62.86,76.85)(0.22,0.12){2}{\line(1,0){0.22}}
\multiput(63.3,77.09)(0.22,0.11){2}{\line(1,0){0.22}}
\multiput(63.74,77.32)(0.22,0.11){2}{\line(1,0){0.22}}
\multiput(64.18,77.54)(0.23,0.11){2}{\line(1,0){0.23}}
\multiput(64.64,77.75)(0.23,0.1){2}{\line(1,0){0.23}}
\multiput(65.09,77.95)(0.23,0.1){2}{\line(1,0){0.23}}
\multiput(65.55,78.14)(0.23,0.09){2}{\line(1,0){0.23}}
\multiput(66.01,78.33)(0.47,0.17){1}{\line(1,0){0.47}}
\multiput(66.48,78.5)(0.47,0.16){1}{\line(1,0){0.47}}
\multiput(66.94,78.67)(0.47,0.16){1}{\line(1,0){0.47}}
\multiput(67.42,78.82)(0.48,0.15){1}{\line(1,0){0.48}}
\multiput(67.89,78.97)(0.48,0.14){1}{\line(1,0){0.48}}
\multiput(68.37,79.1)(0.48,0.13){1}{\line(1,0){0.48}}
\multiput(68.85,79.23)(0.48,0.12){1}{\line(1,0){0.48}}
\multiput(69.33,79.35)(0.49,0.11){1}{\line(1,0){0.49}}
\multiput(69.82,79.46)(0.49,0.1){1}{\line(1,0){0.49}}
\multiput(70.31,79.56)(0.49,0.09){1}{\line(1,0){0.49}}
\multiput(70.79,79.64)(0.49,0.08){1}{\line(1,0){0.49}}
\multiput(71.29,79.72)(0.49,0.07){1}{\line(1,0){0.49}}
\multiput(71.78,79.79)(0.49,0.06){1}{\line(1,0){0.49}}
\multiput(72.27,79.85)(0.49,0.05){1}{\line(1,0){0.49}}
\multiput(72.77,79.9)(0.5,0.04){1}{\line(1,0){0.5}}
\multiput(73.26,79.94)(0.5,0.03){1}{\line(1,0){0.5}}
\multiput(73.76,79.97)(0.5,0.02){1}{\line(1,0){0.5}}
\multiput(74.25,79.99)(0.5,0.01){1}{\line(1,0){0.5}}
\put(74.75,80){\line(1,0){0.5}}
\multiput(75.25,80)(0.5,-0.01){1}{\line(1,0){0.5}}
\multiput(75.75,79.99)(0.5,-0.02){1}{\line(1,0){0.5}}
\multiput(76.24,79.97)(0.5,-0.03){1}{\line(1,0){0.5}}
\multiput(76.74,79.94)(0.5,-0.04){1}{\line(1,0){0.5}}
\multiput(77.23,79.9)(0.49,-0.05){1}{\line(1,0){0.49}}
\multiput(77.73,79.85)(0.49,-0.06){1}{\line(1,0){0.49}}
\multiput(78.22,79.79)(0.49,-0.07){1}{\line(1,0){0.49}}
\multiput(78.71,79.72)(0.49,-0.08){1}{\line(1,0){0.49}}
\multiput(79.21,79.64)(0.49,-0.09){1}{\line(1,0){0.49}}
\multiput(79.69,79.56)(0.49,-0.1){1}{\line(1,0){0.49}}
\multiput(80.18,79.46)(0.49,-0.11){1}{\line(1,0){0.49}}
\multiput(80.67,79.35)(0.48,-0.12){1}{\line(1,0){0.48}}
\multiput(81.15,79.23)(0.48,-0.13){1}{\line(1,0){0.48}}
\multiput(81.63,79.1)(0.48,-0.14){1}{\line(1,0){0.48}}
\multiput(82.11,78.97)(0.48,-0.15){1}{\line(1,0){0.48}}
\multiput(82.58,78.82)(0.47,-0.16){1}{\line(1,0){0.47}}
\multiput(83.06,78.67)(0.47,-0.16){1}{\line(1,0){0.47}}
\multiput(83.52,78.5)(0.47,-0.17){1}{\line(1,0){0.47}}
\multiput(83.99,78.33)(0.23,-0.09){2}{\line(1,0){0.23}}
\multiput(84.45,78.14)(0.23,-0.1){2}{\line(1,0){0.23}}
\multiput(84.91,77.95)(0.23,-0.1){2}{\line(1,0){0.23}}
\multiput(85.36,77.75)(0.23,-0.11){2}{\line(1,0){0.23}}
\multiput(85.82,77.54)(0.22,-0.11){2}{\line(1,0){0.22}}
\multiput(86.26,77.32)(0.22,-0.11){2}{\line(1,0){0.22}}
\multiput(86.7,77.09)(0.22,-0.12){2}{\line(1,0){0.22}}
\multiput(87.14,76.85)(0.22,-0.12){2}{\line(1,0){0.22}}
\multiput(87.57,76.61)(0.21,-0.13){2}{\line(1,0){0.21}}
\multiput(88,76.35)(0.21,-0.13){2}{\line(1,0){0.21}}
\multiput(88.42,76.09)(0.21,-0.14){2}{\line(1,0){0.21}}
\multiput(88.84,75.82)(0.21,-0.14){2}{\line(1,0){0.21}}
\multiput(89.25,75.54)(0.2,-0.14){2}{\line(1,0){0.2}}
\multiput(89.65,75.25)(0.2,-0.15){2}{\line(1,0){0.2}}
\multiput(90.05,74.96)(0.13,-0.1){3}{\line(1,0){0.13}}
\multiput(90.45,74.66)(0.13,-0.1){3}{\line(1,0){0.13}}
\multiput(90.84,74.34)(0.13,-0.11){3}{\line(1,0){0.13}}
\multiput(91.22,74.03)(0.13,-0.11){3}{\line(1,0){0.13}}
\multiput(91.59,73.7)(0.12,-0.11){3}{\line(1,0){0.12}}
\multiput(91.96,73.37)(0.12,-0.11){3}{\line(1,0){0.12}}
\multiput(92.32,73.03)(0.12,-0.12){3}{\line(1,0){0.12}}
\multiput(92.68,72.68)(0.12,-0.12){3}{\line(0,-1){0.12}}
\multiput(93.03,72.32)(0.11,-0.12){3}{\line(0,-1){0.12}}
\multiput(93.37,71.96)(0.11,-0.12){3}{\line(0,-1){0.12}}
\multiput(93.7,71.59)(0.11,-0.13){3}{\line(0,-1){0.13}}
\multiput(94.03,71.22)(0.11,-0.13){3}{\line(0,-1){0.13}}
\multiput(94.34,70.84)(0.1,-0.13){3}{\line(0,-1){0.13}}
\multiput(94.66,70.45)(0.1,-0.13){3}{\line(0,-1){0.13}}
\multiput(94.96,70.05)(0.15,-0.2){2}{\line(0,-1){0.2}}
\multiput(95.25,69.65)(0.14,-0.2){2}{\line(0,-1){0.2}}
\multiput(95.54,69.25)(0.14,-0.21){2}{\line(0,-1){0.21}}
\multiput(95.82,68.84)(0.14,-0.21){2}{\line(0,-1){0.21}}
\multiput(96.09,68.42)(0.13,-0.21){2}{\line(0,-1){0.21}}
\multiput(96.35,68)(0.13,-0.21){2}{\line(0,-1){0.21}}
\multiput(96.61,67.57)(0.12,-0.22){2}{\line(0,-1){0.22}}
\multiput(96.85,67.14)(0.12,-0.22){2}{\line(0,-1){0.22}}
\multiput(97.09,66.7)(0.11,-0.22){2}{\line(0,-1){0.22}}
\multiput(97.32,66.26)(0.11,-0.22){2}{\line(0,-1){0.22}}
\multiput(97.54,65.82)(0.11,-0.23){2}{\line(0,-1){0.23}}
\multiput(97.75,65.36)(0.1,-0.23){2}{\line(0,-1){0.23}}
\multiput(97.95,64.91)(0.1,-0.23){2}{\line(0,-1){0.23}}
\multiput(98.14,64.45)(0.09,-0.23){2}{\line(0,-1){0.23}}
\multiput(98.33,63.99)(0.17,-0.47){1}{\line(0,-1){0.47}}
\multiput(98.5,63.52)(0.16,-0.47){1}{\line(0,-1){0.47}}
\multiput(98.67,63.06)(0.16,-0.47){1}{\line(0,-1){0.47}}
\multiput(98.82,62.58)(0.15,-0.48){1}{\line(0,-1){0.48}}
\multiput(98.97,62.11)(0.14,-0.48){1}{\line(0,-1){0.48}}
\multiput(99.1,61.63)(0.13,-0.48){1}{\line(0,-1){0.48}}
\multiput(99.23,61.15)(0.12,-0.48){1}{\line(0,-1){0.48}}
\multiput(99.35,60.67)(0.11,-0.49){1}{\line(0,-1){0.49}}
\multiput(99.46,60.18)(0.1,-0.49){1}{\line(0,-1){0.49}}
\multiput(99.56,59.69)(0.09,-0.49){1}{\line(0,-1){0.49}}
\multiput(99.64,59.21)(0.08,-0.49){1}{\line(0,-1){0.49}}
\multiput(99.72,58.71)(0.07,-0.49){1}{\line(0,-1){0.49}}
\multiput(99.79,58.22)(0.06,-0.49){1}{\line(0,-1){0.49}}
\multiput(99.85,57.73)(0.05,-0.49){1}{\line(0,-1){0.49}}
\multiput(99.9,57.23)(0.04,-0.5){1}{\line(0,-1){0.5}}
\multiput(99.94,56.74)(0.03,-0.5){1}{\line(0,-1){0.5}}
\multiput(99.97,56.24)(0.02,-0.5){1}{\line(0,-1){0.5}}
\multiput(99.99,55.75)(0.01,-0.5){1}{\line(0,-1){0.5}}

\linethickness{0.3mm}
\multiput(9.38,44.38)(0.21,0.12){214}{\line(1,0){0.21}}
\linethickness{0.3mm}
\multiput(9.38,65.62)(0.21,-0.12){214}{\line(1,0){0.21}}
\linethickness{0.3mm}
\qbezier(90,75)(89.54,75.8)(83.15,81.94)
\qbezier(83.15,81.94)(76.76,88.08)(70,90)
\qbezier(70,90)(56.74,92.07)(43.94,87.52)
\qbezier(43.94,87.52)(31.13,82.96)(20,75)
\qbezier(20,75)(14.44,71.32)(9.86,66.32)
\qbezier(9.86,66.32)(5.28,61.33)(5,55)
\qbezier(5,55)(5.28,48.67)(9.86,43.68)
\qbezier(9.86,43.68)(14.44,38.68)(20,35)
\qbezier(20,35)(31.13,27.04)(43.94,22.48)
\qbezier(43.94,22.48)(56.74,17.93)(70,20)
\qbezier(70,20)(76.76,21.92)(83.15,28.06)
\qbezier(83.15,28.06)(89.54,34.2)(90,35)
\linethickness{0.3mm}
\put(10,15.62){\line(1,0){39.38}}
\put(49.38,15.62){\vector(1,0){0.12}}
\put(19.38,21.88){\makebox(0,0)[cc]{Flow}}

\put(72.5,56.25){\makebox(0,0)[cc]{$A$}}

\put(34.38,71.88){\makebox(0,0)[cc]{$W_N$}}

\put(34.38,35.88){\makebox(0,0)[cc]{$W_S$}}

\put(13,54){\makebox(0,0)[cc]{$W_B$}}

\put(42.5,54){\makebox(0,0)[cc]{$W_L$}}

\end{picture}
\caption{The warning zones for an island}
\label{figwarning}
\end{center}
\vspace{-0.15in}
\end{figure}

Consider a ship moving from the West (left) to the East (right) of the diagram in Figure~\ref{figwarning}, and trying to avoid the circular island labelled $A$. The boat is carried by the flow of the water, 1 m/s in an Easterly direction, and it also has an onboard motor, which has 3 settings, 1 m/s North relative to the flow, 1 m/s South or off. Periodically the boat measures its position to within a certain error.

\noindent

We give a simple algorithm: The boat sails straight East unless it enters a warning zone $W$ around an island. The shape of the warning zone is modified to allow for possible error in determining the boat's position. The warning zone $W$ is split into 4 parts or \textit{modes}, $W_N$, $W_S$, $W_B$ and $W_L$ so $W =  W_N \cup W_S \cup W_B \cup W_L$, as show in in Figure~\ref{figwarning}. 
If the boat's position is in $W_N$, it steers North to avoid the island, and if it is in $W_S$, it steers South. If it is in $W_B$, it can choose to steer either North or South to avoid the island (e.g., by tossing a coin).  If it is in $W_L$ it cannot avoid hitting the island. 

If the position is checked often enough and the warning zones are drawn correctly, the boat can always avoid hitting a single island. However, if the boat has large errors, or delays in a measurement, or should there be two nearby islands, things might be very different.

\section{Specification and verification: Preconditions and postconditions for modes} \label{versec7}

A strategy is a plan of action designed to achieve a long-term or overall aim.  We shall interpret a strategy as a list of conditional statements of the form: if something is true then choose to do a certain action. To verify a system requires a verification of a strategy, for which there are two problems: 

1.  the strategy is incomplete in the sense that we find the system in a position where the strategy does not give an option, and 

2.  the strategy is unsuccessful in the sense that applying it to the system does not achieve the aims.

The existence of modes and the necessity of changing modes leads to the idea of a strategy that involves a means of choosing new modes.

\subsection{The local world of a mode $\alpha$} \label{worldalpha}

In Section \ref{specyy}, we gave the safety and liveness problem we want to solve and outlined a theoretical framework involving measurement and computation. Our approach to the control of physical systems by separating the physical behaviour into distinct modes implies that a path from start $S$ to end $E$ may cross any mode many times or not at all.

\begin{figure}[htbp]
\begin{center}
\unitlength 0.4 mm
\begin{picture}(135,77)(10,23)
\linethickness{0.3mm}
\qbezier(20,60)(19.94,52.21)(24.75,45.59)
\qbezier(24.75,45.59)(29.56,38.98)(40,32.5)
\qbezier(40,32.5)(50.32,25.97)(64.16,23.56)
\qbezier(64.16,23.56)(77.99,21.16)(97.5,22.5)
\qbezier(97.5,22.5)(117.08,23.72)(125.5,30.94)
\qbezier(125.5,30.94)(133.92,38.16)(132.5,52.5)
\qbezier(132.5,52.5)(131.28,66.84)(124.06,74.66)
\qbezier(124.06,74.66)(116.84,82.48)(102.5,85)
\qbezier(102.5,85)(88.19,87.61)(78.56,88.81)
\qbezier(78.56,88.81)(68.94,90.02)(62.5,90)
\qbezier(62.5,90)(56.02,90.02)(49.41,88.22)
\qbezier(49.41,88.22)(42.79,86.41)(35,82.5)
\qbezier(35,82.5)(27.17,78.63)(23.56,73.22)
\qbezier(23.56,73.22)(19.95,67.8)(20,60)
\linethickness{0.3mm}
\qbezier(43.12,86.25)(51.9,72.54)(58.37,66.83)
\qbezier(58.37,66.83)(64.83,61.11)(70,62.5)
\qbezier(70,62.5)(75.2,63.77)(79.41,66.78)
\qbezier(79.41,66.78)(83.62,69.79)(87.5,75)
\qbezier(87.5,75)(91.43,80.21)(92.18,83.07)
\qbezier(92.18,83.07)(92.93,85.93)(90.62,86.88)
\linethickness{0.3mm}
\qbezier(20,60)(33.03,57.41)(40.25,54.41)
\qbezier(40.25,54.41)(47.47,51.4)(50,47.5)
\qbezier(50,47.5)(52.62,43.62)(52.62,38.81)
\qbezier(52.62,38.81)(52.62,34)(50,27.5)
\linethickness{0.3mm}
\qbezier(130,65)(109.11,57.2)(100.69,51.19)
\qbezier(100.69,51.19)(92.27,45.17)(95,40)
\qbezier(95,40)(97.56,34.79)(102.52,31.34)
\qbezier(102.52,31.34)(107.49,27.88)(115.62,25.62)
\linethickness{0.3mm}
\qbezier(40,50)(55.62,47.39)(65.84,46.19)
\qbezier(65.84,46.19)(76.07,44.98)(82.5,45)
\qbezier(82.5,45)(88.97,44.99)(95.74,46.04)
\qbezier(95.74,46.04)(102.51,47.09)(110.62,49.38)
\linethickness{0.3mm}
\qbezier(40,45)(50.4,43.05)(58.22,41.92)
\qbezier(58.22,41.92)(66.04,40.79)(72.5,40.31)
\qbezier(72.5,40.31)(78.95,39.81)(87.97,40.93)
\qbezier(87.97,40.93)(96.99,42.06)(110,45)
\linethickness{0.3mm}
\put(112.49,46.63){\line(0,1){0.49}}
\multiput(112.39,46.15)(0.09,0.48){1}{\line(0,1){0.48}}
\multiput(112.21,45.71)(0.09,0.22){2}{\line(0,1){0.22}}
\multiput(111.95,45.33)(0.13,0.19){2}{\line(0,1){0.19}}
\multiput(111.62,45.02)(0.11,0.1){3}{\line(1,0){0.11}}
\multiput(111.24,44.81)(0.19,0.11){2}{\line(1,0){0.19}}
\multiput(110.83,44.7)(0.41,0.11){1}{\line(1,0){0.41}}
\put(110.42,44.7){\line(1,0){0.42}}
\multiput(110.01,44.81)(0.41,-0.11){1}{\line(1,0){0.41}}
\multiput(109.63,45.02)(0.19,-0.11){2}{\line(1,0){0.19}}
\multiput(109.3,45.33)(0.11,-0.1){3}{\line(1,0){0.11}}
\multiput(109.04,45.71)(0.13,-0.19){2}{\line(0,-1){0.19}}
\multiput(108.86,46.15)(0.09,-0.22){2}{\line(0,-1){0.22}}
\multiput(108.76,46.63)(0.09,-0.48){1}{\line(0,-1){0.48}}
\put(108.76,46.63){\line(0,1){0.49}}
\multiput(108.76,47.12)(0.09,0.48){1}{\line(0,1){0.48}}
\multiput(108.86,47.6)(0.09,0.22){2}{\line(0,1){0.22}}
\multiput(109.04,48.04)(0.13,0.19){2}{\line(0,1){0.19}}
\multiput(109.3,48.42)(0.11,0.1){3}{\line(1,0){0.11}}
\multiput(109.63,48.73)(0.19,0.11){2}{\line(1,0){0.19}}
\multiput(110.01,48.94)(0.41,0.11){1}{\line(1,0){0.41}}
\put(110.42,49.05){\line(1,0){0.42}}
\multiput(110.83,49.05)(0.41,-0.11){1}{\line(1,0){0.41}}
\multiput(111.24,48.94)(0.19,-0.11){2}{\line(1,0){0.19}}
\multiput(111.62,48.73)(0.11,-0.1){3}{\line(1,0){0.11}}
\multiput(111.95,48.42)(0.13,-0.19){2}{\line(0,-1){0.19}}
\multiput(112.21,48.04)(0.09,-0.22){2}{\line(0,-1){0.22}}
\multiput(112.39,47.6)(0.09,-0.48){1}{\line(0,-1){0.48}}

\linethickness{0.3mm}
\put(42.18,47.25){\line(0,1){0.49}}
\multiput(42.09,46.77)(0.08,0.48){1}{\line(0,1){0.48}}
\multiput(41.93,46.32)(0.16,0.45){1}{\line(0,1){0.45}}
\multiput(41.69,45.91)(0.12,0.2){2}{\line(0,1){0.2}}
\multiput(41.39,45.57)(0.1,0.12){3}{\line(0,1){0.12}}
\multiput(41.03,45.3)(0.18,0.14){2}{\line(1,0){0.18}}
\multiput(40.63,45.11)(0.2,0.09){2}{\line(1,0){0.2}}
\multiput(40.21,45.01)(0.42,0.1){1}{\line(1,0){0.42}}
\put(39.79,45.01){\line(1,0){0.43}}
\multiput(39.37,45.11)(0.42,-0.1){1}{\line(1,0){0.42}}
\multiput(38.97,45.3)(0.2,-0.09){2}{\line(1,0){0.2}}
\multiput(38.61,45.57)(0.18,-0.14){2}{\line(1,0){0.18}}
\multiput(38.31,45.91)(0.1,-0.12){3}{\line(0,-1){0.12}}
\multiput(38.07,46.32)(0.12,-0.2){2}{\line(0,-1){0.2}}
\multiput(37.91,46.77)(0.16,-0.45){1}{\line(0,-1){0.45}}
\multiput(37.82,47.25)(0.08,-0.48){1}{\line(0,-1){0.48}}
\put(37.82,47.25){\line(0,1){0.49}}
\multiput(37.82,47.75)(0.08,0.48){1}{\line(0,1){0.48}}
\multiput(37.91,48.23)(0.16,0.45){1}{\line(0,1){0.45}}
\multiput(38.07,48.68)(0.12,0.2){2}{\line(0,1){0.2}}
\multiput(38.31,49.09)(0.1,0.12){3}{\line(0,1){0.12}}
\multiput(38.61,49.43)(0.18,0.14){2}{\line(1,0){0.18}}
\multiput(38.97,49.7)(0.2,0.09){2}{\line(1,0){0.2}}
\multiput(39.37,49.89)(0.42,0.1){1}{\line(1,0){0.42}}
\put(39.79,49.99){\line(1,0){0.43}}
\multiput(40.21,49.99)(0.42,-0.1){1}{\line(1,0){0.42}}
\multiput(40.63,49.89)(0.2,-0.09){2}{\line(1,0){0.2}}
\multiput(41.03,49.7)(0.18,-0.14){2}{\line(1,0){0.18}}
\multiput(41.39,49.43)(0.1,-0.12){3}{\line(0,-1){0.12}}
\multiput(41.69,49.09)(0.12,-0.2){2}{\line(0,-1){0.2}}
\multiput(41.93,48.68)(0.16,-0.45){1}{\line(0,-1){0.45}}
\multiput(42.09,48.23)(0.08,-0.48){1}{\line(0,-1){0.48}}

\put(75,52){\makebox(0,0)[cc]{tube}}

\put(57.5,76.88){\makebox(0,0)[cc]{}}

\put(69,75){\makebox(0,0)[cc]{$A_\alpha$}}
\put(63,97){\makebox(0,0)[cc]{avoidance zone}}

\put(106.88,73.12){\makebox(0,0)[cc]{$\alpha$}}

\put(41.5,40.5){\makebox(0,0)[cc]{$S_{(\alpha,i)}$}}

\put(115,36.88){\makebox(0,0)[cc]{$E_{(\alpha,i)}$}}

\put(7,33){\makebox(0,0)[cc]{$\alpha$ entry zone}}

\put(154,33){\makebox(0,0)[cc]{$\alpha$ exit zone}}

%\put(84.38,91.88){\makebox(0,0)[cc]{avoidance zone}}

\end{picture}
\caption{A safety and liveness problem in mode $\alpha$}
\label{figlocalsafe}
\end{center}
\vspace{-0.15in} 
\end{figure}
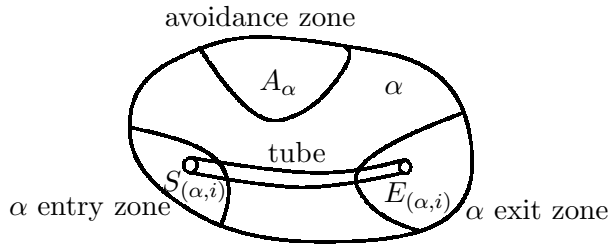

For each mode $\alpha$
we must specify an entry zone  $S_{(\alpha,i)}  \subset X$ and exit zone $E_{(\alpha,i)} \subset X$ local to $\alpha$ --- these are distinct from the start and end of the original global problem.
In Figure~\ref{figlocalsafe} we look at a local version of Figure~\ref{figtube} on a mode $\alpha$, where the local entry and exit zones are indexed by an integer $i$, as we may have several of these for the mode $\alpha$. Typically the avoidance zone $A_\alpha$ in mode $\alpha$ will include the intersection of the avoidance zone $A$ for the whole system with the subset of the state space $X_{\alpha}$ corresponding to $\alpha$, but may be larger due to problems with measurements etc.

In programming, a pair consisting of a \textit{precondition} $p$ and \textit{postcondition} $q$ is essentially a guarantee about the behaviour of a program $S$. In the classic notation of Floyd-Hoare triples for programs, $\{p\}S\{q\}$ asserts that on executing program $S$ on all input states satisfying precondition $p$, on termination of $S$, the output final states satisfies postcondition $q$. This idea has several generalisations and roles depending on the nature of the conditions $p$ and $q$ and the termination of $S$.  One purpose of triples is to give someone who is not familiar with the workings of the program $S$ enough information to correctly integrate it with another program in a manner allowing, or requiring, ignorance of the implementation. Pre and postconditions can be composed if the processes they belong to can be composed: a postcondition for the first process can imply the precondition for the second: 
\begin{center}
$\{p\}S\{q'\}$ and $\{p'\}S\{q\}$  and $q' \Longrightarrow p'$   implies   $\{p\}S\{q\}$. 
\end{center}
Forms of information hiding have become prominent as modularity has become pervasive through object oriented programming. 

We shall extend this technique to our class of analogue-digital dynamical systems, where the job of measuring or calculating things to do with the physical system has been subcontracted to the modes, and it is at that level that we must implement the preconditions and postconditions, interpreting the zones $S_{(\alpha,i)}$ as a precondition and $E_{(\alpha,i)}$ as a postcondition. 

What information is needed to specify the pre and post conditions? For algorithms and programs the answer is simple. The states and behaviour of the program is defined by the semantics of the programming language. Furthermore, since we are thinking about running code, the state and behaviour is further determined by properties of the language implementation and the computer system. The essential point is that all the information that is relevant, or can be known, lies within the software and the application problem. 
Thus the control system's recognition of the pre and post conditions can only take place in $\mathrm{State}_\alpha$, the dataset describing the system in mode $\alpha$, \textit{not} the data set $X$ in the real world system. Also the information about the pre and post conditions themselves is specified in $\mathrm{State}_\alpha$. 

We have a precondition
$\mathrm{pre}_{(\alpha,i)} \subset \mathrm{State}_\alpha$ whose relation to the state space is: If the data describing the system is in $\mathrm{pre}_{(\alpha,i)} \subset \mathrm{State}_\alpha$ then the system is in a state in $S_{(\alpha,i)} \subset X$. 
Similarly, we have a postcondition $\mathrm{post}_{(\alpha,i)} \subset \mathrm{State}_\alpha$ whose relation is: If the system is in a state in $E_{(\alpha,i)} \subset X$ then data describing the system is in $\mathrm{post}_{(\alpha,i)} \subset \mathrm{State}_\alpha$. In addition we need a set of control parameters 
$\mathrm{orders}_{(\alpha,i)} \subset \mathrm{State}_\alpha$ such that the following is true: 

\begin{guarantee}
For the system controlled by mode $\alpha$'s control algorithm with control parameters $\mathrm{orders}_{(\alpha,i)}$, if mode $\alpha$'s digital image of the system (i.e.,\ an element of $\mathrm{State}_\alpha$) is in the subset $\mathrm{pre}_{(\alpha,i)} \subset \mathrm{State}_\alpha$  at some timestep, then at some future timestep it will be in $\mathrm{post}_{(\alpha,i)} \subset \mathrm{State}_\alpha$. 
\end{guarantee}

This guarantee may seem rather vague on the idea of time, but of course time coordinates (as measured by some clock used as a sensor) could easily be part of the information specified in the pre or post conditions. 

\subsection{Changing modes} \label{cvow}

\begin{figure}[htbp]
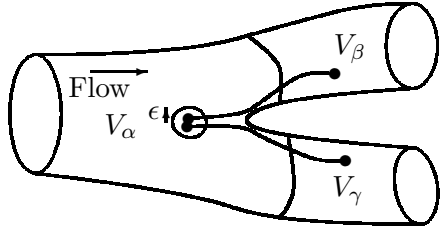

\begin{center}
\unitlength 0.45 mm

\caption{Measurement error gives a nondeterministic system}
\label{figbifyy}
\end{center}
\vspace{-0.15in}
\end{figure}

We specify the modes by taking a collection of known pre and post conditions for each mode. From the point of view of controlling of the actual behaviour of a system, we are not necessarily dealing with a deterministic system because of the nature of measurements and their interpretation. Consider a state space of the form of Figure~\ref{figbifyy}. From the first position it is not possible to say which fork the flow will lead to. (This is just another form of the island avoidance problem in Figure~\ref{figwarning}.) Thus, in changing modes, in general, there may be several preconditions belonging to different new modes that could hold for a given postcondition, and we may have to choose which precondition to follow. 

\begin{assumption} \label{trias}
Suppose every mode $\alpha\in\mathcal{M}$ possess a finite set $\mathcal{T}_\alpha$ of triples, which we index by $i$, having the form
\[
(\mathrm{pre}_{(\alpha,i)},\mathrm{orders}_{(\alpha,i)},\mathrm{post}_{(\alpha,i)})
\]
Here $\mathrm{pre}_{(\alpha,i)}\subset \mathrm{State}_\alpha$ is a precondition such that on entering $\alpha$ the behaviour evolves and enters the postcondition $\mathrm{post}_{(\alpha,i)}\subset \mathrm{State}_\alpha$, under the assumption that mode $\alpha$ is given the control parameters $\mathrm{orders}_{(\alpha,i)}$. We will assume that all these triples are safe in the sense that the postcondition includes the statement that the avoidance zone has been avoided while the system has evolved from the precondition to the postcondition. We shall often refer to the triple above as $(\alpha,i)$.  Let $\mathcal{T}$ be the set of all triples for all modes, i.e., $\cup_{\beta\in\mathcal{M}}\mathcal{T}_\beta$.

\end{assumption}

We need to know what happens to these subsets under change of mode.

\begin{definition}
For a set $D\subset X_\beta$ we specify a set $D(\alpha\to\beta)\subset \mathrm{State}_\alpha$ as the set of elements in $ \mathrm{State}_\alpha$ which end up in $D\subset X_\beta$ under a change of mode from $\alpha$ to $\beta$. 
\end{definition}

If we were in a postcondition $\mathrm{post}_{(\alpha,i)} \subset \mathrm{State}_\alpha$ and had a single precondition $\mathrm{pre}_{(\beta,k)} \subset X_\beta$ such that $\mathrm{post}_{(\alpha,i)} \subset \mathrm{pre}_{(\beta,k)} (\alpha\to \beta)$, then all we have to do is to change mode to $\beta$, then we will be in $\mathrm{pre}_{(\beta,k)} \subset X_\beta$ and by applying control parameters $\mathrm{orders}_{(\beta,k)}$ we continue on the chain of pre and post conditions. But, as noted above, in many cases the best we can do is to have a \textit{choice} of new preconditions, 
and to make the choice we use selection functions.

\begin{definition} \label{arrdefpre}
Given subsets $Y$ and $Z_j$, for $j$ in a finite collection $\mathcal{J}$, of $\mathrm{State}_\alpha$ we say that the collection
$Z_j$ \textit{is compatible with or follows from} $Y$, if we have computable functions $\phi_j:\mathrm{State}_\alpha\to \{0,1\}$ for $j\in\mathcal{J}$, called \textit{selection functions}, satisfying the following conditions:

1. If $\phi_j(x)=1$ then $x\in Z_j$.

2. If $x\in Y$ then there is at least one $j\in\mathcal{J}$ so that $\phi_j(x)=1$.

\noindent For this we use the notation
\[
Y \Subset \bigcup_{j\in\mathcal{J}} Z_j\ .
\]
\end{definition}

\begin{definition} \label{strategy}
Let $\mathcal{T}$ be the set of all triples for all modes of a system. 
A strategy will be a choice of triples $\mathcal{C}_{(\alpha,i)} \subset \mathcal{T}$ for every $(\alpha,i)\in \mathcal{T}$. Using the power set $PF(\mathcal{T})$ of finite subsets, a strategy is represented by a function $\mathcal{C}:\mathcal{T}\to PF(\mathcal{T})$. 
\end{definition}

\begin{definition} \label{arrdef}
A strategy $\mathcal{C}:\mathcal{T}\to P(\mathcal{T})$ is said to be complete at $(\alpha,i)\in \mathcal{T}$ if 
\[
\mathrm{post}_{(\alpha,i)}  \Subset \bigcup_{(\beta,j)\in\mathcal{C}_{(\alpha,i)}} \mathrm{pre}_{(\beta,j)} (\alpha\to\beta)
\]
\end{definition}

Now supposing that the strategy is complete at the triple $(\alpha,i)\in \mathcal{T}$ we consider how to pass control from the mode and orders
$(\alpha,\mathrm{orders}_{(\alpha,i)})$ to another mode satisfying a precondition. We use the selection functions $\phi_{(\beta,j)}$ as in Definition~\ref{arrdefpre}.

\medskip

\textbf{Procedure: transfer control from $(\alpha,i)$} 

\textbf{input} $n_{\mathrm{max}}$

\textbf{for} $n=0$ \textbf{to} $n_{\mathrm{max}}$ \textbf{do}

\quad \textbf{run control algorithm $(\alpha,i)$ for one time step to update $x\in\mathrm{state}_\alpha$} 

\quad\textbf{for} $(\beta,j)\in\mathcal{C}_{(\alpha,i)}$ \textbf{do}

\qquad\textbf{if} $\phi_{(\beta,j)}(x)=1$ \textbf{then transfer control to} $(\beta,j)$

\quad \textbf{rof}

\textbf{rof}

\medskip
If the precondition for $(\alpha,i)$ was satisfied, then at some point we shall find $\phi_{(\beta,j)}(x)=1$ and transfer to another triple.
Note that we have a search of an unordered set, as all we want is to find {\em{any}} $\phi_{(\beta,j)}(x)=1$, as we have not ranked them by preference. The selection functions $\phi_{(\beta,j)}$ need to be computable to be a component of the algorithm.

\subsection{A verification algorithm for strategies} \label{verity}
To allow a uniform description of the verification algorithm, it will be convenient to describe the start and end zones of the safety and liveness problem of Section~\ref{specyy} in terms of pre and post-conditions. 
We take a triple \textit{start} for the Start Zone which automatically has the postcondition satisfied (so we can immediately transfer out of it) and a triple \textit{end} for the End Zone which automatically has the precondition satisfied (so we can freely transfer into it). 
In fact, as we only have a detailed description of the state on each mode, we may have to have several \textit{start} triples if one mode cannot describe all of the Start Zone, and likewise for the End Zone, but that is not a serious complication.

For a given strategy $\mathcal{C}:\mathcal{T}\to P(\mathcal{T})$ we begin by constructing a \textit{strategy graph} with vertices labelled by the triples.
If the strategy is complete at $(\alpha,i)\in \mathcal{T}$ then we draw arrows from $(\alpha,i)$ to every triple in $\mathcal{C}_{(\alpha,i)}$. If the strategy is not complete at $(\alpha,i)\in \mathcal{T}$ then we draw no arrows from $(\alpha,i)$. This can be thought of as having arrows for guaranteed exits, and no arrows if we have no guarantees.
 Now we follow every possible path along the arrows from every \textit{start} triple. If all of these paths reach an \textit{end} triple, then the strategy is \textit{verified}, and repeated use of the above \textbf{transfer control from $(\alpha,i)$} algorithm will get to the finish.

\begin{example}
An example of a verified strategy would be Figure~\ref{graffin}, where we omit any triples not involved in the path from \textit{start}:

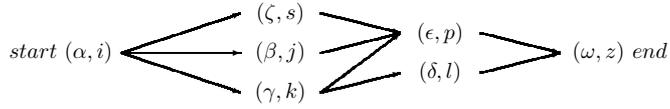
\begin{figure}[htbp]
\begin{center}
\unitlength 0.7 mm
\scalebox{.75}{
\begin{picture}(130,35)(10,50)
\put(5,65){\makebox(0,0)[cc]{$start\ (\alpha,i)$}}

\put(60,75){\makebox(0,0)[cc]{$(\zeta,s)$}}
\put(60,65){\makebox(0,0)[cc]{$(\beta,j)$}}
\put(60,55){\makebox(0,0)[cc]{$(\gamma,k)$}}
\put(100,70){\makebox(0,0)[cc]{$(\epsilon,p)$}}

\put(100,60){\makebox(0,0)[cc]{$(\delta,l)$}}

\put(145,65){\makebox(0,0)[cc]{$(\omega,z)\ end$}}

\linethickness{0.3mm}
\multiput(20,65)(0.36,0.12){83}{\line(1,0){0.36}}
\put(50,75){\vector(3,1){0.12}}
\linethickness{0.3mm}
\put(20,65){\line(1,0){30}}
\put(50,65){\vector(1,0){0.12}}
\linethickness{0.3mm}
\multiput(20,65)(0.36,-0.12){83}{\line(1,0){0.36}}
\put(50,55){\vector(3,-1){0.12}}
\linethickness{0.3mm}
\multiput(70,75)(0.48,-0.12){42}{\line(1,0){0.48}}
\put(90,70){\vector(4,-1){0.12}}
\linethickness{0.3mm}
\multiput(70,65)(0.48,0.12){42}{\line(1,0){0.48}}
\put(90,70){\vector(4,1){0.12}}
\linethickness{0.3mm}
\multiput(70,55)(0.48,0.12){42}{\line(1,0){0.48}}
\put(90,60){\vector(4,1){0.12}}
\linethickness{0.3mm}
\multiput(110,70)(0.48,-0.12){42}{\line(1,0){0.48}}
\put(130,65){\vector(4,-1){0.12}}
\linethickness{0.3mm}
\multiput(110,60)(0.48,0.12){42}{\line(1,0){0.48}}
\put(130,65){\vector(4,1){0.12}}
\linethickness{0.3mm}
\multiput(70,55)(0.16,0.12){125}{\line(1,0){0.16}}
\put(90,70){\vector(4,3){0.12}}
\end{picture}
}
\caption{A strategy graph for verification}
\label{graffin}
\end{center}
\vspace{-0.15in}
\end{figure}

\noindent
To construct this graph we would have to find selection functions $f_{(\zeta,s)},f_{(\beta,j)},f_{(\gamma,k)}: \mathrm{State}_\alpha \to\{0,1\}$ showing that
\begin{eqnarray*}
\mathrm{post}_{(\alpha,i)} & \Subset&  \mathrm{pre}_{(\zeta,s)} (\alpha\to\zeta)\, \cup \, \mathrm{pre}_{(\beta,j)} (\alpha\to\beta) \cr &&\cup\,
\mathrm{pre}_{(\gamma,k)} (\alpha\to\gamma) 
\end{eqnarray*}
for the leftmost vertex, and then similarly for the other vertices except the rightmost.
\end{example}

Note that it makes the process of verification more likely to work
if we choose smaller sets $\mathcal{C}_{(\alpha,i)}$ satisfying the condition in Definition~\ref{arrdef} rather than larger ones. (This just prevents the verification  process worrying about possibilities which we can arrange not to happen.) However, there is not necessarily a unique choice of smallest $\mathcal{C}_{(\alpha,i)}$. Most obviously there may be many choices of triples with different control parameters but the same preconditions. 
The task of finding a strategy (choice of $\mathcal{C}$) which verifies, as opposed to verifying a single given strategy, is much longer. 

\subsection{The limits of verification}

{ \quote
\textit{No plan survives contact with the enemy.} 
\vspace{-0.15in}
\begin{flushright}
After Helmuth von Moltke the Elder (1800-1891)
\end{flushright}
\endquote}
\vspace{-0.15in}

A successful verification of the sort described in Section~\ref{verity} is a design milestone, but it is neither a necessary nor sufficient condition for solving our safety and liveness problem.
For computationally controlling a real world physical system, we are primarily interested in ensuring the physical system does what it should, but understanding what exactly the physical system should do in all sorts of circumstances is a major specification problem. 

A final verification process is  built solely around a software specification, which is an independent  abstraction removed from the  the real world. To make our point we consider four complex control systems where the design and implementation was carried out to the highest possible standard -- as were the investigations into the failure of the real world systems. We see a wide variety of causes and comment on how a possible verification of the control system would have influenced the outcome. 

\medskip
\noindent
1. \textit{A `larger than expected' measurement}. Two quotes from the Mars probe Schiaparelli anomaly inquiry (\cite{Schiaparelli} p.\ 12), with abbreviations expanded in square brackets:
\begin{quote}
- The parachute was deployed, and the parachute inflation triggered some oscillations of Schiaparelli at a frequency of approximately 2.5 Hz.

- About 0.2 sec after the peak load of the parachute inflation, the IMU [Inertial Measurement Unit] measured a pitch angular rate (angular rate around Z-EDM axis) larger than expected.

- The IMU raised a saturation flag,.
\end{quote}

\noindent
and later on the same page (underline in original document)

\begin{quote}
Because of the error in the estimated attitude that occurred at parachute inflation, the GNC [Guidance Navigation and Control] Software projected the RDA range measurements with an erroneous off-vertical angle and deduced a \ul{negative altitude (cosinus of angles $>$ 90 degrees are negative)}. There was no check on board of the plausibility of this altitude calculation
\end{quote}

\noindent
Consequently the parachute was cut free prematurely, and the Mars probe crashed. In principle, if a software specification did not allow for a `larger than expected' measurement, then there is no reason why a software verification process would pick up a problem.
\newline
\newline
\noindent
2. \textit{Incompleteness of the model}. The most probable cause of the Mars Polar Lander crash was triggering a magnetic sensor on deploying landing legs being misinterpreted as landing and shutting down the descent engine (\cite{Polar} p.\ 26). This problem does not lie in a component, but with the \textit{incompleteness of the model}. When the legs were extending, the sensor triggered in a manner which has been assumed only to occur at landing. No amount of verification from a software specification would pick up this problem, only a detailed theoretical analysis or testing in the field would do it. 
\newline
\newline
\noindent
3. \textit{Software exception?} It might be said that the destruction of the Ariane 501 launcher was due to an out of range exception on type conversion, but that over-simplifies the picture given by the Inquiry Board report, from which we quote part of the chain of technical events in reverse time (\cite{Ariane5} p.\ 3-4):

\begin{quote}
$\bullet$ The reason why the active SRI 2 [Inertial Reference System] did not send correct attitude data was that the unit had declared a failure due to a software exception.

$\bullet$ The OBC [On-Board Computer] could not switch to the back-up SRI 1 because that unit had already ceased to function during the previous data cycle (72 milliseconds period) for the same reason as SRI 2.

$\bullet$ The internal SRI software exception was caused during execution of a data conversion from 64-bit floating point to 16-bit signed integer value. The floating point number which was converted had a value greater than what could be represented by a 16-bit signed integer. This resulted in an Operand Error. The data conversion instructions (in Ada code) were not protected from causing an Operand Error, although other conversions of comparable variables in the same place in the code were protected.

$\bullet$ The error occurred in a part of the software that only performs alignment of the strap-down inertial platform. This software module computes meaningful results only before lift-off. As soon as the launcher lifts off, this function serves no purpose.

$\bullet$ The alignment function is operative for 50 seconds after starting of the Flight Mode of the SRIs which occurs at H0 - 3 seconds for Ariane 5. Consequently, when lift-off occurs, the function continues for approx. 40 seconds of flight. This time sequence is based on a requirement of Ariane 4 and is not required for Ariane 5.

$\bullet$ The Operand Error occurred due to an unexpected high value of an internal alignment function result called BH, Horizontal Bias, related to the horizontal velocity sensed by the platform. This value is calculated as an indicator for alignment precision over time.

$\bullet$ The value of BH was much higher than expected because the early part of the trajectory of Ariane 5 differs from that of Ariane 4 and results in considerably higher horizontal velocity values.
\end{quote}

In the third item we see the software exception, together with a lack of protection which one might imagine would be found by some software verification process. However the last two items contain another reason -- a change in the real world system which was not reflected in the software. To quote the report again: `The design of the Ariane 5 SRI is practically the same as that of an SRI which is presently used on Ariane 4, particularly as regards the software.'
Only if this new range of values was built into a software specification could a software verification find the problem.
In the fourth item we see the irony that the error in was in a software module whose output was only required at an earlier stage in the flight, so the error could have simply been ignored. 
\newline
\newline
\noindent
4. \textit{Inconsistency in measurements}. A quote from the final report on the airliner crash of 1st.\ June 2009 in the Atlantic (\cite{aircrash2009}, Synposis p.\ 17): 

\begin{quote}

At 2 h 10 min 05, likely following the obstruction of the Pitot probes by ice crystals, the speed indications were incorrect and some automatic systems disconnected. The aeroplane's flight path was not controlled by the two copilots. They were rejoined 1 minute 30 later by the Captain, while the aeroplane was in a stall situation that lasted until the impact with the sea at 2 h 14 min 28.

The accident resulted from the following succession of events:

$\bullet$ˆ Temporary inconsistency between the measured airspeeds, likely following the obstruction of the Pitot probes by ice crystals that led in particular to autopilot disconnection and a reconfiguration to alternate law, [List continues...]

\end{quote}

\noindent
Unless a mechanism for dealing with an inconsistency in measurement was written into a software specification, a software verification would not pick up any problem. 

\smallskip

We may  ask, with 20:20 hindsight, what general design lessons might be drawn from these cases. \textit{Each type of problem can be viewed as a mode transition problem.} 

\smallskip
Both Mars crashes (1,2) were likely caused by an premature transition between modes, in these cases between parachute descent, powered descent, and landing. From a mode-orientated design approach attention would be focused on mode transitions, and we might ask just what sensors on the probes could provide additional information on the timing of this mode transition. In particular we are not simply talking about adding 
redundancy by replication of existing sensors (pointless in case (2)) but about integrating as much different information as possible. 

In case (3) we might ask if dividing the flight into modes might reduce the complexity of the system at any one time, making it easier to avoid unnecessary software modules at a given stage, and easier to update when the real world effectively changes. In particular, there should be an {\em{encapsulation}} property that the control software for one mode should not be allowed to intervene in the operation of another mode, and as this can be explicitly stated in a software specification it could, in principle, be verified. In our current environment, the necessity to harden systems against cyber attack makes such encapsulation even more important.

In case (4) we can ask if there is a way to safely avoid autopilot disconnection despite inconsistent data. Note that an automatic stall warning did occur (\cite{aircrash2009} p.\ 22) and that a stall is a completely different mode from normal flight, requiring different control decisions. Of course, a decision to disconnect is itself a mode transition in the wider physical system context, and since it is that real world system which is important, not the software, the consequences of such a mode transition would need to be studied in detail.

\subsection{Beyond certainty}
In general, it should be obvious that the more factors we take into account, and the more model or component failures we allow for, the more difficult the verification will become, until eventually it is likely to be impossible to verify a complex system. Nonetheless, this is a better prospect than verifying an incomplete system as 

(1) it makes the assumptions necessary for verification explicit and allows a possibility of estimating the chance of failure (allowing a choice of \textit{safest} route), and 

(2) by identifying the likely failure modes we can try to minimise their impact, including 

(3) implementing control strategies which only have a chance of avoiding failure but are better than nothing.

\section{Physical modes as charts for manifolds} \label{secman}

\subsection{Measuring the state}

Now we return to the measurement of the state problem in Section~\ref{modmes}. We assume that we have a finite number of sensors which report real values (or sequences of reals). For any particular position $z\in X$ the values reported by some sensors will be invalid (e.g.\ `out of range') but if we get enough valid reports we should be able to estimate $z\in X$. As the results are real numbers, we end up plotting the position of  $z\in X$ in some $\mathbb{R}^n$. As we add more results $n$ increases, but the redundancy of the results leads to the measured values for points of $X$ in a neighbourhood of $z\in X$ clustering near certain subsets of $\mathbb{R}^n$ defined by relations among the measured values.\footnote{Consider the measurement problem of finding a mobile phone in a country by triangulation from transmitters. For various reasons (e.g., out of range) many transmitters will report that the phone cannot be seen. If at least three transmitters give a valid distance we can estimate a position. If more give a valid response, we can make the position more accurate, or even diagnose problems with the transmitters.}.  

In many physical systems it is common to write equations of motion as differential equations. If we put this together with the idea of open sets of $X$ being represented by coordinates in $\mathbb{R}^n$ then it might be natural to think of $X$ as some sort of manifold. 
%However, this is certainly not always the case: For example 
%it takes many fewer coordinates to describe the motion of an aircraft carrier with its aircraft on board than it takes to describe the carrier and planes when the planes have been launched -- these modes naturally have a different dimension. Nonetheless it will be instructive to consider the case of motion on a manifold. 

It may be tempting to try to describe dynamics on a manifold by some global description, such as imbedding the manifold in $\mathbb{R}^n$. However, in practice for many systems which are already complicated this only adds more complexity, and it is much easier to restrict to coordinate charts and pay the price of transition between charts (e.g.\ \cite{AdCaHiMa}).

\subsection{Dynamics on a chart in a manifold}
We consider the case where the state space $X$ is a compact manifold. Essentially an $m$-dimensional manifold is a topological space which locally looks like an open subset of $\mathbb{R}^m$, or more precisely (e.g., see \cite{LeeManif}) we have

\begin{definition} A m-dimensional manifold is a paracompact Hausdorff topological space $X$ that has

(i) an index set $\mathcal{M}$

(i) a collection of open sets $V_\alpha\subset X$ for $\alpha\in\mathcal{M}$, called \textit{charts}, which has union $X$. 

(ii) homeomorphisms $\phi_\alpha:V_\alpha\to U_\alpha$, where $U_\alpha$ is an open subset of $\mathbb{R}^m$. 

(iii)  invertible smooth (i.e.\ arbitrarily differentiable) maps (called transition functions) between the following open subsets of  $\mathbb{R}^m$ defined by:
\[
\tau_{\beta\alpha}=\phi_\beta\,\phi_\alpha{}^{-1}: \phi_\alpha(V_\alpha\cap V_\beta) \to \phi_\beta(V_\alpha\cap V_\beta)\ .
\]
%which obey the rules that $\tau_{\alpha\alpha}$ the identity and
%\[
%\tau_{\beta\gamma}\circ \tau_{\gamma\alpha}=\tau_{\beta\alpha}:\phi_\alpha(V_\alpha\cap V_\beta\cap V_\gamma) \to \phi_\beta(V_\alpha\cap V_\beta\cap V_\gamma)\ .
%\]
\end{definition}

\begin{figure}[htbp]
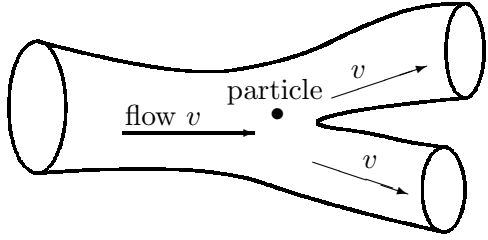

\begin{center}
\unitlength 0.5 mm

\caption{A vector field $v$ on a part of a manifold}
\label{figpant}
\end{center}
\vspace{-0.15in}
\end{figure}

Figure~\ref{figpant} shows a part of a manifold and a vector field, with dynamics given by flow along the vector field. In fact at the beginning of our example in Section~\ref{manpre} we already considered the local picture of a dynamical system on a coordinate chart 
$U_\alpha\subset \mathbb{R}^n$. 
The equation of motion was given by
(\ref{eqrn}) for the vector field $v=(v_1,\dots,v_n)$ (a function of the coordinates and possibly time) and the control
$c=(c_1,\dots,c_n)$ (specified by an algorithm). Also in Figure~\ref{figlocalsafe} we considered the local version of the safety and liveness problem pictured in Figure~\ref{figtube}.

For convenience we continue to suppose that there is a single metric specified on the whole manifold. 
We take $X$ to be a Riemannian manifold \cite{RiemGeom,RindEssRel}, which gives a uniform idea of distance which is not dependent on which coordinate chart we use. 
 This means that the distance along a path where $x=(x_1,\dots,x_n)$ is a specified function of the variable $s$ from $s_0$ to $s_1$ is defined to be the following, where $g_{ij}$ are functions of position:
\begin{eqnarray} \label{riedist}
\int_{s=s_0}^{s_1} \sqrt{\sum_{i,j}  g_{ij}\, \frac{ \mathrm{d} x_i }{\mathrm{d} s}\,   \frac{ \mathrm{d} x_j }{\mathrm{d} s}    }\ \mathrm{d} s
\end{eqnarray}

\subsection{Measurement, prediction and control on a chart} \label{meprcochart}
Now we reconsider Section~\ref{meprco} in the case of operating on a chart $U_\alpha\subset \mathbb{R}^n$ where the dynamics are given by the differential equation (\ref{eqrn}) in Section~\ref{manpre}. For a fixed choice of control parameters this differential equation has known coefficients. 
In this chart we examine the $(\lambda,\epsilon,\eta)$-property of Definition~\ref{vermot}. The first thing is to examine the available sensors (assuming that a sufficient number of these are operating) and determine the accuracy $\epsilon$ to which position may be determined. Then we integrate the differential equation forward for a time step $\lambda$ using some numerical method, e.g.\ Runge-Kutta. Now we see what the error $\eta$ of the estimated position is. There are three sources of error, and we label them according to the type of data numbered in Section~\ref{vcgaksuy}; (2) the error of measurement, (3) the error of the model (in this case a differential equation) in modelling reality and (4) the error of the numerical method for the given computational time. It is very likely that to get a reasonable 
$(\lambda,\epsilon,\eta)$-property we shall have to restrict to some bounded subset of $U_\alpha\subset \mathbb{R}^n$, but this is fine as long as the union of the `shrunk' charts still cover $X$, and is a standard mathematical procedure. 

We are now in a position to try to solve the local version of the safety and liveness problem Figure~\ref{figtube}, given start $S_{(\alpha,i)}$ and end $E_{(\alpha,i)}$ regions for a choice of control parameters $\mathrm{orders}_{(\alpha,i)}$. (For the moment we just assume that these are specified, we will address this later.) However we need to be careful about the avoidance zone $A_\alpha$. There will likely be `bad' values of $x\in U_\alpha$ for 
which an $\epsilon$ neighbourhood of $x$ under the one step path prediction algorithm has corresponding $\eta$ neighbourhood not a subset of the chart - i.e.\ points which have a chance of leaving the chart in one time step. Even if the differential equation looks perfectly nice, in principle we have no idea what real world states correspond to elements of $\mathbb{R}^n$ which are not in $U_\alpha$, and such dubious calculations should only be used in control systems in a case of utter desperation. Setting $B_{(\alpha,i)}$ to be this set of bad values, we require that 
$B_{(\alpha,i)}$ does not intersect $S_{(\alpha,i)}$, as it is not good to have a precondition including possibilities where we lose control of the system in one time step. Now set $A_\alpha=(A\cap U_\alpha)\cup (B_{(\alpha,i)}\setminus E_{(\alpha,i)})$, where $A\cap U_\alpha$ is the relevant part of the original avoidance zone, and $B_{(\alpha,i)}\setminus E_{(\alpha,i)}$ forbids the system to enter the region $B_{(\alpha,i)}$ where it can leave the chart in one time step, unless the end condition $E_{(\alpha,i)}$ is valid. 

It is now a probably lengthy calculation to check if all paths beginning on $S_{(\alpha,i)}$ end in $E_{(\alpha,i)}$ without entering $A_\alpha$ -- and in this case $(S_{(\alpha,i)}, \mathrm{orders}_{(\alpha,i)}, E_{(\alpha,i)})$ for a triple in the sense of Assumption~\ref{trias} in Section~\ref{cvow}. At the end of each time step we end in a radius $\eta$ neighbourhood, and the next timestep has to consider a sufficient number of $\epsilon$ neighbourhoods to cover this $\eta$ neighbourhood -- if the system has high dimension this could be a large number. The regions mentioned above need to be specified in a computable manner, for example a computable function which is 1 on $A_\alpha$ and is zero for all points more than $\epsilon$ from $A_\alpha$. 
The final stage in the analysis is to calculate the inverse images of the preconditions under the change of coordinate maps, construct the partition of unity for Definition~\ref{arrdef}, and construct the graphs for the verification process. 

The analysis presented here is in some sense a worst case scenario -- if there are explicit solutions to the equations of motion or if a lot is known about the solutions it may not be too difficult to check the pre and post conditions case by case, helped by using longer (maybe only one) timesteps in the calculation. (In general one of the advantages of the modes is to reduce the complexity of the local systems, making reasoning about them easier.) This brings us to the point about the choices involved or orders and possible pre and post conditions: a priori finding these seems like looking for a needle in a haystack. In practice we require an ability to reason about the system to see what sort of pre or post conditions are likely to occur. As these choices are required at the design stage we can look to the reasoning ability of the system designers to provide them. The key stage is often identifying an approximation to the model, and then refining the approximation. 
It should be pointed out that progress is being made on formal methods for reasoning about physical systems (e.g.\ \cite{BrownKleerConf}), so one day it may be practical for machines to analyse and design systems on their own, or to exercise their judgement in that most human activity of `making up things as we go along'.

\section{Case Study: Autonomous Racing Cars} \label{carstt}
\begin{figure}[htbp]
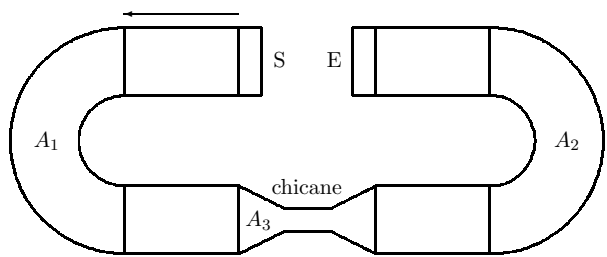

\begin{center}
\unitlength 0.8 mm
\scalebox{.75}{

}
\caption{A track for racing cars, marking the spatial areas $A_1,A_2,A_3\subset [0,L]$ that are the components of the avoidance zone}
\label{chic}
\end{center}
\vspace{-0.15in}
\end{figure}

Consider a racing track of length $L$ with two cars and a chicane, a narrow region of track where the cars might collide as in Figure~\ref{chic}. The cars (labelled 1 and 2) have to go from the start line $S$ to the finish line $E$ taking care in the potential danger areas $A_1$, $A_2$ and $A_3$. The aim is to do this in a short time, and we consider an autonomous control system for each car.

\subsection{The state space and physical behaviour}

We assume that the cars cannot go backwards, that their maximum speed is $120\,Km/hr$, and that $L$ is the length of the track. 
We take the state space to consist of 4-tuples
$(x_1,v_1,x_2,v_2)$ where for $i=1,2$ $x_i\in [0,L]$ is the position of car $i$ and $v_i\in [0,120]$ is its speed.  Thus, the state space for a single car is $[0,L]\times [0,120]$, and for both cars, taking car 1 first, is 
\[
X=[0,L]\times [0,120] \times [0,L]\times [0,120]\ .
\]

The behaviour of the system is given by $f:[0,\infty)\to X$, which gives the actual state of both of the cars at time $t$. We suppose that the positions $x_1$ and $x_2$ of cars 1,2 respectively are measured within an error of 1$\, m$. The avoidance zone can be summarised by the conditions
\begin{compactitem}
\item Take the bends slowly: For car $i$ at the bends, that is $x_i\in A_1\cup A_2$, we must avoid $v_i >80$, where $80\,Km/hr$ is the maximum safe speed to avoid coming off the track at the bends. 
\item Can't have both cars on the chicane at the same time: The cars each have their own lanes, and do not normally have a chance to collide. But at the chicane the track narrows, and to avoid collision we do not want both $x_1$ and $x_2$ in $A_3$. 
\end{compactitem}
The avoidance zone $A$ as a subspace of the state space $X$ is
\begin{eqnarray*}
&& A = A_3 \times [0,120] \times A_3 \times [0,120] \cr
&&\qquad \bigcup \   (A_1\cup A_2)\times (80,120] \times [0,L] \times [0,120] \cr
&&\qquad  \bigcup \,  [0,L] \times [0,120]  \times (A_1\cup A_2)\times (80,120] \ .
\end{eqnarray*}

Note that the state space could be considered as a manifold (technically with boundary and corners as we have taken closed intervals), but 
we avoid the complication of transition functions by taking a single coordinate system for all of the state space.

\subsection{Assigning modes}
Suppose that we control car 1 (there is only one difference between the cars, noted later).
Divide the operation of the system into modes for car 1:
\[
\mathcal{M}=\{ \mathrm{Sta},\mathrm{Str},\mathrm{Ben},\mathrm{Ch.gw},\mathrm{Ch.rt},\mathrm{End} \}
\]
which in more detail are

\smallskip

\noindent $\mathrm{Sta}$\quad \textbf{Start:}\ Accelerating from the start. 

\noindent $\mathrm{Str}$\quad  \textbf{Straight:}\ Racing down a straight piece of track

\noindent $\mathrm{Ben}$\quad \textbf{Bend:}\ Taking care not to come off the track at a bend

\noindent $\mathrm{Ch.gw}$\quad \textbf{Chicane: give way}\ Stop to let the other car enter the chicane first

\noindent $\mathrm{Ch.rt}$\quad \textbf{Chicane: race through:}\ Race through the chicane

\noindent $\mathrm{End}$\quad \textbf{End:}\ Crossing the finishing line and stopping.

\begin{figure}[htbp]
\begin{center}
\unitlength 0.7 mm
\scalebox{.75}{
\begin{picture}(110.62,59)(10,30)
\linethickness{0.3mm}
\put(60,40){\line(1,0){45}}
\linethickness{0.3mm}
\put(105,40){\line(0,1){40}}
\linethickness{0.3mm}
\multiput(60,40)(0.14,0.12){333}{\line(1,0){0.14}}
\linethickness{0.3mm}
\put(10,40){\line(1,0){50}}
\linethickness{0.3mm}
\multiput(10,80)(0.15,-0.12){333}{\line(1,0){0.15}}
\linethickness{0.3mm}
\put(60,40){\line(0,1){40}}
\linethickness{0.3mm}
\multiput(95.62,71.88)(0.12,-0.12){78}{\line(1,0){0.12}}
\linethickness{0.3mm}
\multiput(90.62,66.88)(0.12,-0.12){120}{\line(1,0){0.12}}
\linethickness{0.3mm}
\multiput(85,62.5)(0.12,-0.12){167}{\line(1,0){0.12}}
\linethickness{0.3mm}
\multiput(80,57.5)(0.12,-0.12){146}{\line(1,0){0.12}}
\linethickness{0.3mm}
\multiput(100,75)(0.12,-0.12){42}{\line(1,0){0.12}}
\linethickness{0.3mm}
\multiput(75,52.5)(0.12,-0.12){104}{\line(1,0){0.12}}
\linethickness{0.3mm}
\multiput(70.62,49.38)(0.12,-0.12){78}{\line(1,0){0.12}}
\linethickness{0.3mm}
\multiput(65.62,44.38)(0.12,-0.12){36}{\line(1,0){0.12}}
\put(56.25,35){\makebox(0,0)[cc]{Straight}}

\put(115,84){\makebox(0,0)[cc]{Chicane: give way}}

\put(110.62,35){\makebox(0,0)[cc]{Chicane: race through}}

\put(61,84){\makebox(0,0)[cc]{Start}}

\put(13,84){\makebox(0,0)[cc]{End}}

\put(13.12,35){\makebox(0,0)[cc]{Bend}}

\put(10,79.5){\makebox(0,0)[cc]{$\bullet$}}
\put(60,79.5){\makebox(0,0)[cc]{$\bullet$}}
\put(105.8,79.4){\makebox(0,0)[cc]{$\bullet$}}

\put(10,39.7){\makebox(0,0)[cc]{$\bullet$}}
\put(60,39.7){\makebox(0,0)[cc]{$\bullet$}}
\put(105,39.7){\makebox(0,0)[cc]{$\bullet$}}

% add points and labels

%\put(60,70){\makebox(0,0)[cc]{$\bullet$}}
%\put(95,65){\makebox(0,0)[cc]{$\bullet$}}
%
%\put(64,70){\makebox(0,0)[cc]{$A$}}
%\put(99,65){\makebox(0,0)[cc]{$B$}}

\end{picture}
}
\caption{The graph of mode transitions for car 1}
\label{fignervetm}
\end{center}
\vspace{-0.15in}
\end{figure}
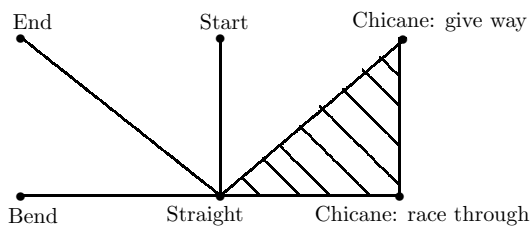

\smallskip
The possible transitions between these modes are shown in Figure~\ref{fignervetm}. The modes are the dots, and a line between modes indicates that a transition between those modes is possible. In most of the cases on this diagram, these are simple transitions, with only one choice. However the shaded triangle between the straight and the two chicane modes indicates a contested choice: before entering the chicane we need to choose between braking and accelerating, and the choice may not be obvious. 

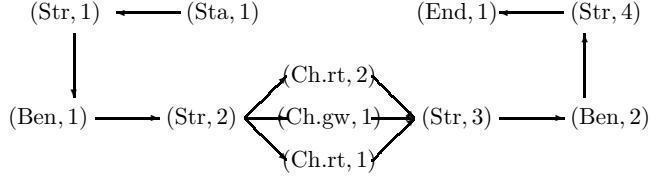
\begin{figure}[htbp]

\unitlength 0.7 mm

\begin{center}
\scalebox{.8}{
\begin{picture}(130,40)(5,30)
\linethickness{0.3mm}
\put(10,45){\line(0,1){15}}
\put(10,45){\vector(0,-1){0.12}}
\linethickness{0.3mm}
\put(15,40){\line(1,0){15}}
\put(30,40){\vector(1,0){0.12}}
\linethickness{0.3mm}
\put(20,65){\line(1,0){15}}
\put(20,65){\vector(-1,0){0.12}}
\put(45,65){\makebox(0,0)[cc]{$(\mathrm{Sta},1)$}}

\put(8,65){\makebox(0,0)[cc]{$(\mathrm{Str},1)$}}

\put(4,40){\makebox(0,0)[cc]{$(\mathrm{Ben},1)$}}

\put(40,40){\makebox(0,0)[cc]{$(\mathrm{Str},2)$}}
\put(70,40){\makebox(0,0)[cc]{$(\mathrm{Ch.gw},1)$}}
\put(70,30){\makebox(0,0)[cc]{$(\mathrm{Ch.rt},1)$}}
\put(70,50){\makebox(0,0)[cc]{$(\mathrm{Ch.rt},2)$}}

\linethickness{0.3mm}
\multiput(50,40)(0.12,0.12){83}{\line(1,0){0.12}}
\put(60,50){\vector(1,1){0.12}}
\linethickness{0.3mm}
\put(50,40){\line(1,0){10}}
\put(60,40){\vector(1,0){0.12}}
\linethickness{0.3mm}
\multiput(50,40)(0.12,-0.12){83}{\line(1,0){0.12}}
\put(60,30){\vector(1,-1){0.12}}
\linethickness{0.3mm}
\multiput(80,50)(0.12,-0.12){83}{\line(1,0){0.12}}
\put(90,40){\vector(1,-1){0.12}}
\linethickness{0.3mm}
\put(80,40){\line(1,0){10}}
\put(90,40){\vector(1,0){0.12}}
\linethickness{0.3mm}
\multiput(80,30)(0.12,0.12){83}{\line(1,0){0.12}}
\put(90,40){\vector(1,1){0.12}}
\linethickness{0.3mm}
\put(110,40){\line(1,0){15}}
\put(125,40){\vector(1,0){0.12}}
\put(100,40){\makebox(0,0)[cc]{$(\mathrm{Str},3)$}}
\put(136,40){\makebox(0,0)[cc]{$(\mathrm{Ben},2)$}}
\put(135,65){\makebox(0,0)[cc]{$(\mathrm{Str},4)$}}
\put(100,65){\makebox(0,0)[cc]{$(\mathrm{End},1)$}}

\linethickness{0.3mm}
\put(130,45){\line(0,1){15}}
\put(130,60){\vector(0,1){0.12}}
\linethickness{0.3mm}
\put(110,65){\line(1,0){15}}
\put(110,65){\vector(-1,0){0.12}}
\end{picture}
}

\caption{The strategy graph for car 1}
\label{fignervet}
\vspace{-0.15in}
\end{center}

\end{figure}

As in Section~\ref{versec7} we consider the set $\mathcal{T}$ of triples of preconditions, orders and postconditions. We need only one $(\mathrm{Sta},1)$ for the start and one $(\mathrm{End},1)$ for the end. However it is convenient to have four triples $(\mathrm{Str},1)$ to $(\mathrm{Str},4)$ for the straights (labelled in ascending order from the start), as they all have different lengths and different recommended exit speeds. Similarly, we have two $(\mathrm{Ben},1)$ and $(\mathrm{Ben},2)$ for the two bends. The complication comes at the chicane, where we already have two modes, and it is convenient to take two triples $(\mathrm{Ch.rt},1)$ and $(\mathrm{Ch.rt},2)$ for the race through case, depending on whether the other car is in front or behind. 
Following Figure~\ref{graffin} in Section~\ref{verity}, we would like to have our version of the strategy graph looking like
Figure~\ref{fignervet}.

We have the problem of getting the pre and post conditions to match. We shall ignore the simpler cases and go to the chicane, where we have to specify the modes $\mathrm{Ch.rt},\mathrm{Ch.gw}$ and the triples
$(\mathrm{Ch.rt},1)$, $(\mathrm{Ch.rt},2)$ and $(\mathrm{Ch.gw},1)$, together with the selection functions (see Definition~\ref{arrdefpre}) demonstrating that we can actually draw the arrows from $(\mathrm{Str},2)$ in Figure~\ref{fignervet}.

\begin{figure}[htbp]
\begin{center}
\unitlength 0.6 mm
\scalebox{.75}{
\begin{picture}(110,85)(-20,0)
\linethickness{0.5mm}
\put(30,60){\line(0,1){20}}
\linethickness{0.5mm}
\put(30,60){\line(1,0){30}}
\linethickness{0.5mm}
\put(30,30){\line(1,0){30}}
\linethickness{0.5mm}
\put(60,10){\line(0,1){20}}
\linethickness{0.5mm}
\put(25,65){\line(1,0){40}}
\linethickness{0.5mm}
\put(65,25){\line(0,1){40}}
\linethickness{0.5mm}
\put(25,25){\line(1,0){40}}
\linethickness{0.5mm}
\put(25,25){\line(0,1){40}}
\linethickness{0.1mm}
\multiput(30,70)(0.12,-0.12){83}{\line(1,0){0.12}}
\linethickness{0.1mm}
\multiput(30,80)(0.12,-0.12){167}{\line(1,0){0.12}}
\linethickness{0.1mm}
\multiput(40,80)(0.12,-0.12){167}{\line(1,0){0.12}}
\linethickness{0.1mm}
\multiput(50,80)(0.12,-0.12){83}{\line(1,0){0.12}}
\linethickness{0.1mm}
\multiput(50,30)(0.12,-0.12){83}{\line(1,0){0.12}}
\linethickness{0.1mm}
\multiput(40,30)(0.12,-0.12){167}{\line(1,0){0.12}}
\linethickness{0.1mm}
\multiput(30,30)(0.12,-0.12){167}{\line(1,0){0.12}}
\linethickness{0.1mm}
\multiput(30,20)(0.12,-0.12){83}{\line(1,0){0.12}}
\linethickness{0.1mm}
\multiput(25,60)(0.12,0.12){42}{\line(1,0){0.12}}
\linethickness{0.1mm}
\multiput(25,50)(0.12,0.12){125}{\line(1,0){0.12}}
\linethickness{0.1mm}
\multiput(25,40)(0.12,0.12){208}{\line(1,0){0.12}}
\linethickness{0.1mm}
\multiput(25,30)(0.12,0.12){292}{\line(1,0){0.12}}
\linethickness{0.1mm}
\multiput(30,25)(0.12,0.12){292}{\line(1,0){0.12}}
\linethickness{0.1mm}
\multiput(40,25)(0.12,0.12){208}{\line(1,0){0.12}}
\linethickness{0.1mm}
\multiput(50,25)(0.12,0.12){125}{\line(1,0){0.12}}
\linethickness{0.1mm}
\multiput(60,25)(0.12,0.12){42}{\line(1,0){0.12}}
\linethickness{0.5mm}
\put(35,5){\line(0,1){80}}
\linethickness{0.5mm}
\put(55,5){\line(0,1){80}}
\linethickness{0.1mm}
\put(35,85){\line(1,0){20}}
\linethickness{0.1mm}
\put(35,80){\line(1,0){20}}
\linethickness{0.1mm}
\put(35,75){\line(1,0){20}}
\linethickness{0.1mm}
\put(35,70){\line(1,0){20}}
\linethickness{0.1mm}
\put(35,55){\line(1,0){20}}
\linethickness{0.1mm}
\put(35,45){\line(1,0){20}}
\linethickness{0.1mm}
\put(35,50){\line(1,0){20}}
\linethickness{0.1mm}
\put(35,40){\line(1,0){20}}
\linethickness{0.1mm}
\put(35,35){\line(1,0){20}}
\linethickness{0.1mm}
\put(35,20){\line(1,0){20}}
\linethickness{0.1mm}
\put(35,15){\line(1,0){20}}
\linethickness{0.1mm}
\put(35,10){\line(1,0){20}}
\linethickness{0.1mm}
\put(35,5){\line(1,0){20}}
\put(87,70){\makebox(0,0)[cc]{catch up $(\mathrm{Ch.rt},1)$}}

\put(90,17){\makebox(0,0)[cc]{not getting caught}}
\put(90,9){\makebox(0,0)[cc]{$(\mathrm{Ch.rt},2)$}}

\put(94,45){\makebox(0,0)[cc]{give way $(\mathrm{Ch.gw},1)$}}

\put(-25,28){\makebox(0,0)[cc]{car 2 position}}
\put(-25,20){\makebox(0,0)[cc]{plotted vertically}}
\put(-29,8){\makebox(0,0)[cc]{car 1 position}}
\put(-29,0){\makebox(0,0)[cc]{plotted horizontally}}

\put(-90,-23){
\linethickness{0.3mm}
\put(90,20){\line(0,1){35}}
\put(90,55){\vector(0,1){0.12}}
\linethickness{0.3mm}
\put(90,20){\line(1,0){20}}
\put(110,20){\vector(1,0){0.12}}
\put(95,45){\makebox(0,0)[cc]{$x_2$}}
\put(100,25){\makebox(0,0)[cc]{$x_1$}}
}

\linethickness{0.5mm}
\put(60,60){\line(0,1){20}}
\linethickness{0.5mm}
\put(30,10){\line(0,1){20}}

\put(30,0.7){\makebox(0,0)[cc]{$d_1$}}
\put(37,0){\makebox(0,0)[cc]{$e_1$}}
\put(55,0){\makebox(0,0)[cc]{$e_2$}}
\put(62,0.7){\makebox(0,0)[cc]{$d_2$}}

\put(20,25){\makebox(0,0)[cc]{$g_1$}}
\put(20,30){\makebox(0,0)[cc]{$c_1$}}
\put(20,65){\makebox(0,0)[cc]{$g_2$}}
\put(0,59.5){\makebox(0,0)[cc]{$\mathrm{end\_chicane}+2m$}}

\end{picture}
}

\caption{Car 1: The postcondition for Straight $(\mathrm{Str},2)$ (horizontal shading), the preconditions for race through $\mathrm{Ch.rt}$ (NW to SE shading) and the precondition for give way $\mathrm{Ch.gw}$ (SW to NE, just for clarity drawn wider than race through, it should be the same width).}
\label{frectpost5}
\end{center}
\vspace{-0.15in}
\end{figure}
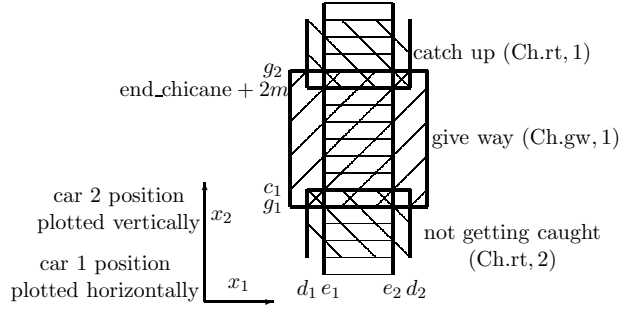

The idea is quite simple: At a certain range of distances $e_1< x_1<e_2$ before the chicane, car 1 has to decide whether to give way $\mathrm{Ch.gw}$ or race through $\mathrm{Ch.rt}$. We take $e_1< x_1<e_2$ to be the postcondition for the previous straight $(\mathrm{Str},2)$, and the distance between $e_1$ and $e_2$ must be large enough to allow measurement and computation of the selection function to take place at the maximum car speed. Race through has two preconditions, one if car 2 is already past the chicane (catch up $(\mathrm{Ch.rt},1)$) and one if car 2 is sufficiently far behind that it is safe to cross the chicane (not getting caught $(\mathrm{Ch.rt},2)$). The give way precondition $(\mathrm{Ch.gw},1)$ is designed to be safe if neither of the other preconditions apply, given that the postcondition for the previous straight $(\mathrm{Str},2)$ is in force. Figure~\ref{frectpost5} shows the car position coordinates $(x_1,x_2)$ for the relevant pre and post conditions.
The overlaps of the preconditions are there to allow for rapid computation of acceptable selection functions, as if we are that close to the boundary between the preconditions, it does not actually matter which one we pick.

We take advantage of the simple dynamics and uniform coordinates to give a rather abbreviated description of the modes for the chicane. We use the variables $\mathrm{begin\_chicane}$ and $\mathrm{end\_chicane}$ to stand for the position coordinates of the beginning and end of the chicane respectively.

\smallskip
\noindent
\textbf{Mode $\mathrm{Ch.rt}$: Chicane: race through} 
\newline\noindent
\textbf{Subset of phase space:} $x_1\in[\mathrm{begin\_chicane}-100m,\mathrm{end\_chicane}+100m]$ 

\medskip
\noindent
\textbf{Mode $\mathrm{Ch.gw}$: Chicane: give way} 
\newline\noindent
\textbf{Subset of phase space:} $x_1\in[\mathrm{begin\_chicane}-100m,\mathrm{end\_chicane}+100m]$ 

\medskip

The triple $(\mathrm{Ch.rt},1)$ is based on the simple idea that if the other car has passed the chicane, it is safe to follow:

\medskip
\noindent
\textbf{Mode: Chicane $\mathrm{Ch.rt}$: race through}  \textit{catch up} for car $1$, label $(\mathrm{Ch.rt},1)$
\newline\noindent
\textbf{Precondition:} $d_1\le x_1\le d_2$ and $x_2\ge \mathrm{end\_chicane}+2m$
\newline\noindent
\textbf{Orders:} Accelerate to $100\ Km/hr$ in 5 seconds and keep going at that speed
\newline\noindent
\textbf{Postcondition:} $x_1\ge \mathrm{end\_chicane}+2m$, $x_2\ge \mathrm{end\_chicane}+2m$, no collision

\medskip

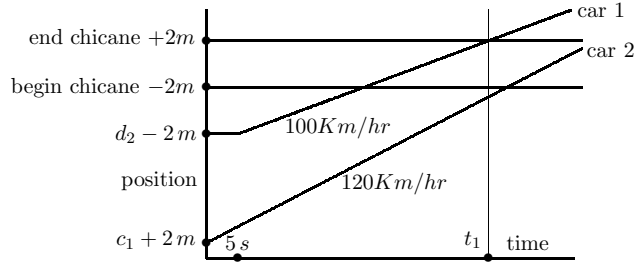
\begin{figure}[htbp]
\begin{center}
\unitlength 0.55 mm
\scalebox{.75}{
\begin{picture}(150,90)(0,-5)
\linethickness{0.3mm}
\put(30,0){\line(1,0){117.5}}
\linethickness{0.3mm}
\put(30,0){\line(0,1){80}}
\linethickness{0.3mm}
\put(30,70){\line(1,0){120}}
\linethickness{0.3mm}
\put(30,55){\line(1,0){120}}
\linethickness{0.3mm}
\put(30,40){\line(1,0){10}}
\linethickness{0.3mm}
\multiput(40,40)(0.32,0.12){333}{\line(1,0){0.32}}
\linethickness{0.3mm}
\multiput(30,5)(0.23,0.12){521}{\line(1,0){0.23}}
\linethickness{0.1mm}
\put(120,0){\line(0,1){80}}
\put(40,0){\makebox(0,0)[cc]{$\bullet$}}

\put(90,24){\makebox(0,0)[cc]{$120Km/hr$}}

\put(72,42){\makebox(0,0)[cc]{$100 Km/hr$}}

\put(30,70){\makebox(0,0)[cc]{$\bullet$}}
\put(0,71){\makebox(0,0)[cc]{end chicane $+2m$}}

\put(30,55){\makebox(0,0)[cc]{$\bullet$}}
\put(-3,55){\makebox(0,0)[cc]{begin chicane $-2m$}}

\put(120,0){\makebox(0,0)[cc]{$\bullet$}}
\put(132.5,5){\makebox(0,0)[cc]{time}}

\put(15,25){\makebox(0,0)[cc]{position}}

\put(30,40){\makebox(0,0)[cc]{$\bullet$}}
\put(14,40){\makebox(0,0)[cc]{$d_2-2\,m$}}

\put(30,5){\makebox(0,0)[cc]{$\bullet$}}
\put(14,6){\makebox(0,0)[cc]{$c_1+2\,m$}}

\put(115,6){\makebox(0,0)[cc]{$t_1$}}

\put(40,5){\makebox(0,0)[cc]{$5\,s$}}

\put(156,80){\makebox(0,0)[cc]{car $1$}}

\put(159,67){\makebox(0,0)[cc]{car $2$}}

\end{picture}
}
\caption{The not getting caught postcondition for \textbf{Chicane: race through}}
\label{figcatch}
\end{center}
\vspace{-0.15in}
\end{figure}

If the other car is sufficiently well behind when our car arrives at the chicane, we have the triple \textit{not getting caught}. Figure~\ref{figcatch} shows the check that needs to be performed to verify that this triple is safe, i.e.\ just what we mean by {\it{sufficiently well behind}} in the form of the constant $c_2$. It compares the fastest motion of car 2 against the slowest motion  of car 1 (according to our model of the cars behaviour and simple dynamics) to ensure that they cannot both cannot be in the chicane at the same time.

\medskip
\noindent
\textbf{Mode $\mathrm{Ch.rt}$: Chicane: race through}  \textit{not getting caught} for car 1,  label $(\mathrm{Ch.rt},2)$
\newline\noindent
\textbf{Precondition:} $d_1\le x_1\le d_2$, $x_2\le c_1$
\newline\noindent
\textbf{Orders:} Accelerate to $100\ Km/hr$ in 5 seconds and keep going at that speed
\newline\noindent
\textbf{Postcondition:} $x_1\ge \mathrm{end\_chicane}+2m$, $x_2\le \mathrm{begin\_chicane}-2m$, no collision

\medskip
\textbf{Chicane: give way} has different triples for the two cars, with the car 1 version following, the car 2 version omits the test $\mathit{timelimit}=0$, so it will wait indefinitely until $x_1\ge \mathrm{end\_chicane}+2m$.

\medskip
\noindent
\textbf{Mode: Chicane: give way}   \textit{wait} for car 1, $(\mathrm{Ch.gw},1)$
\newline\noindent
\textbf{Precondition:} $d_1\le x_1\le d_2$, $g_1\le x_2\le g_2$
\newline\noindent
\textbf{Orders:} 

\quad \textbf{Brake to a halt} 

\quad \textbf{Set} $\mathit{timelimit}=30$ 

\quad \textbf{Repeat} 

\qquad \textbf{Set} $\mathit{timelimit}=\mathit{timelimit}-1$ 

\qquad \textbf{Wait} 1 second

\qquad   \textbf{Oracle call} $x_2:=$ position of car 2

\quad \textbf{Until} 

\qquad $\mathit{timelimit}=0$ \textbf{or}  $x_2\ge \mathrm{end\_chicane}+2m$

\quad \textbf{Accelerate} to $100\ Km/hr$ in 5 seconds and keep going at that speed

\noindent
\textbf{Postcondition:} $x_1\ge  \mathrm{end\_chicane}+2m$ and ($x_2\le \mathrm{begin\_chicane}-2m$ or $x_2\ge \mathrm{end\_chicane}+2m$), no collision

\medskip The position $d_2+1m$ is chosen so that if the car braked to a halt from there then it would come to rest more that $1m$ before the chicane.
The $+2m$ leeway here and elsewhere is because not only the original measurement of the position is subject to a $1m$ error, but also 
we ensure that a subsequent measurement will say that the car is not in the chicane.
We take the 30 second wait time (for a particular race track) so that if car 2 beginning at position $g_1-1m$ were to choose \textbf{Chicane: race through}  then it would have passed the chicane when car $1$'s timer is finished. If car 2 choose \textbf{Mode: Chicane: give way} then as car 2 has no test on $\mathit{timelimit}$ it would not cross into the chicane until car 1 had passed out of the chicane.

%
%*******************
%
%
%
% To make a 
% mathematical model of the chicane it will be convenient to have a picture of the track in Figure~\ref{pulr} with positions $a$, a decision point $b$, the beginning of the danger zone $c$, and a point a short distance after the danger zone $d$. 
%
%\begin{figure}[htbp]
%\begin{center}
%\unitlength 0.6 mm
%\begin{picture}(140,30)(0,35)
%\linethickness{0.3mm}
%\put(10,50){\line(1,0){130}}
%\linethickness{0.3mm}
%\put(70,60){\line(1,0){40}}
%\put(70,45){\line(0,1){15}}
%\put(110,45){\line(0,1){15}}
%\put(70,45){\line(1,0){40}}
%
%\put(20,50){\makebox(0,0)[cc]{$\bullet$}}
%\put(20,55){\makebox(0,0)[cc]{$a$}}
%
%\put(50,50){\makebox(0,0)[cc]{$\bullet$}}
%\put(50,55.6){\makebox(0,0)[cc]{$b$}}
%
%%\put(100,50){\makebox(0,0)[cc]{$\bullet$}}
%%\put(100,55){\makebox(0,0)[cc]{$c$}}
%
%\put(120,50){\makebox(0,0)[cc]{$\bullet$}}
%\put(120,55.6){\makebox(0,0)[cc]{$d$}}
%
%\put(70,50){\makebox(0,0)[cc]{$\bullet$}}
%\put(67,55.6){\makebox(0,0)[cc]{$c$}}
%
%%\put(110,50){\makebox(0,0)[cc]{$\bullet$}}
%%\put(112,55.6){\makebox(0,0)[cc]{$f$}}
%
%\put(90,40){\makebox(0,0)[cc]{chicane danger zone}}
%
%\end{picture}
%\caption{Marked points around the chicane, cars move to the right}
%\label{pulr}
%\end{center}
%\end{figure}
%
% 

From Figure~\ref{frectpost5} we have preconditions for car 1;
\begin{eqnarray*}
\mathrm{pre}_{(\mathrm{Ch.rt},1)} &=& [d_1,d_2] \times [0,120] \times [\mathrm{end\_chicane}+2,L] \cr && \times [0,120]   \cr
\mathrm{pre}_{(\mathrm{Ch.gw},1)} &=& [d_1,d_2] \times [0,120] \times [g_1,g_2] \times [0,120]   \cr
\mathrm{pre}_{(\mathrm{Ch.rt},2)} &=& [d_1,d_2] \times [0,120] \times [0,c_1] \times [0,120]   
\end{eqnarray*}
and we define the selection functions in terms of the characteristic functions $\chi(S)$ of a set
$S\subset X$ (i.e., $\chi(S)(x)$ is 1 if $x\in S$ and zero otherwise). We set $h_1,h_2,k_1,k_2,k_3,k_4$ so that
\begin{eqnarray*}
&&d_1<h_1<e_1<e_2<h_2<d_2 \ \mathrm{and}  \cr
&& g_1<k_1<k_2<c_1<\mathrm{end\_chicane}+2 < k_3<k_4<g_2
\end{eqnarray*}
and then the characteristic functions
\begin{eqnarray*}
\phi_{(\mathrm{Ch.rt},1)} &=& \chi\big([h_1,h_2] \times [0,120] \times [k_3,L] \times [0,120] \big)  \cr
\phi_{(\mathrm{Ch.gw},1)} &=& \chi\big([h_1,h_2] \times [0,120] \times [k_1,k_4] \times [0,120] \big)   \cr
\phi_{(\mathrm{Ch.rt},2)} &=& \chi\big( [h_1,h_2] \times [0,120] \times [0,k_2] \times [0,120]   \big) \ .
\end{eqnarray*}
We see that the condition from Section~\ref{verity} is satisfied, and in the notation used there
\[
\mathrm{post}_{(\mathrm{Str},2)} \Subset \ \mathrm{pre}_{(\mathrm{Ch.rt},1)} \cup 
\mathrm{pre}_{(\mathrm{Ch.gw},1)} \cup 
\mathrm{pre}_{(\mathrm{Ch.rt},2)} \ .
\]
We omit the mode changes as in this simple case all modes use the same coordinates. Of course the $\phi$ are discontinuous functions of variables which are in principle real, and the difficulties of dealing with this are well known \cite{WeiCompAn}. The strict inequalities above allow us to have some degree of error in evaluating these functions, ensuring that it can be done in finite time.

We see that, as far as the algorithm and our model is concerned, we are able to verify the strategy (though we have only considered the chicane, it is the most complicated part). 
Of course our description of the physical system is incomplete, we have not allowed for sensor or actuator (e.g.\ brake) failure. For example, 
car 2 could break down on the chicane while car 1 was running its timer for  \textbf{Chicane: give way}, and then a collision would occur.  However, conveniently omitting these possibilities has allowed us to claim that we have a verified strategy -- not the best way to design a safety critical system.

\section{Concluding remarks}

We have constructed a theoretical framework for analysing certain complex analogue-digital systems whose key features are:

%\begin{enumerate}
\begin{compactitem}
\item an analogue-digital system is viewed as an algorithm with a oracle that is a physical system;
\item in the design of such systems the key feature is data and, primarily, data faithfully representing the actual behaviour of the physical system;
\item the algorithm knows only three types of data:   
\begin{compactitem}
\item data derived from observing and measuring the the actual behaviour;
\item data derived from mathematical models;
\item computable approximations;
\end{compactitem}
\item most real-world examples are complex behaviours that should to be factorised into modes that have independent algorithmic treatments;
\item most real-world examples involve failure modes, i.e., data that derives from behaviours that arise from failures in design or implementation;
\item access to modes can be specified using a modest adaptation of Floyd-Haore triples;
\item key problem is when and how to change from one mode to another;
\end{compactitem}
and specifically, we have shown how 
\begin{compactitem}
\item metric spaces provide a general platform for discussing the four distinct types of data in Section~\ref{vcgaksuy};
\item manifolds with their charts provide a model for both data and modes.
\end{compactitem}
%\end{enumerate}

It is interesting to explore further the theoretical framework. The process of changing modes needs to be analysed and modelled more deeply. The discrete nature of the mode selection (Section \ref{cvow}) needs to be replaced by a continuous model in tune with all the data that underpins the choice of modes and final decision. In further work we will introduce a data type called the \textit{nerve} that is simplicial complex. We see mode definition and selection as a complicated topic in its own right, involving advanced constructions and probabilities.

Further case studies will help test and build upon the framework. These need not necessarily be classical control systems of the kinds ubiquitous in science and engineering; they could be human centred systems where the modes represent the activation of different services.

What of autonomy? What of digital twins?  These are two rich ideas in need of definitions and theories. With examples such as driverless vehicles in smart streets in mind, the Autonomous Principle seems to be conceptually pertinent for the first question. For the second question, we could propose a version of the Turing test \cite{Turingtest}: that a control algorithm (connected via suitable interfaces) could not tell the difference between the real world system and the digital model. However, even for a deterministic system, errors in measurement will lead to increasingly divergent behaviour. This would be a dangerous comparison, but in Section~\ref{meprco} we gave a method of quantifying the fit of the digital twin to the real world \cite{GVtwin}.

%
%++++++++++++++++++++++++++++++
%Following our discussion on resources and time for computation, it is interesting to see that in evolutionary biology there is a similar distinction between having behaviours (i.e.\ control parameters) determined genetically and the more realistic position of having algorithms for determining behaviour determined genetically \cite{parentsMcN}. This is essentially related to the ability of a bird to `compute' behaviour, as opposed to taking a theoretical optimal behaviour. 
%++++++++++++++++++++++++++++++
%
%
%

\bibliographystyle{compj}
%\bibliography{ModellingBidders}

\end{document}